\documentclass{aa}  

\usepackage{graphicx}
\usepackage{txfonts}
\usepackage[dvipsnames]{xcolor}
\usepackage[scr]{rsfso}
\usepackage{makecell}
\usepackage{natbib}
\bibpunct{(}{)}{;}{a}{}{,}
\begin{document} 
\newcommand{\red}[1]{\textcolor{red}{#1}}
\newcommand{\blue}[1]{\textcolor{blue}{#1}}
\newcommand{\green}[1]{\textcolor{green}{#1}}
\newcommand*\diff{\mathop{}\!\mathrm{d}}

   \title{The Cosmological analysis of X-ray cluster surveys}

   \subtitle{VII. Bypassing scaling relations with Lagrangian Deep Learning and Simulation-based inference.}
   \author{N. Cerardi
          \inst{1}
          \and
          M. Pierre \inst{2}
          \and
          F. Lanusse \inst{2,3}
          \and
          X. Corap \inst{1}
          }

   \institute{Universit\'e Paris Cit\'e, Universit\'e Paris-Saclay, CEA, CNRS, AIM, F-91191, Gif-sur-Yvette, France
         \and
         Universit\'e Paris-Saclay, Universit\'e Paris Cit\'e, CEA, CNRS, AIM, 91191, Gif-sur-Yvette, France
         \and
         The Flatiron Institute, 162 5th Ave, New York, NY, 10010, USA
             }

   \date{Received XXX; accepted XXX}

 
  \abstract
   {Galaxy clusters, the pinnacle of structure formation in our universe, are a powerful cosmological probe. Several approaches have been proposed to express cluster number counts, but all these methods rely on empirical explicit scaling relations that link observed properties to the total cluster mass. These scaling relations are over-parametrised, inducing some degeneracy with cosmology. Moreover, they do not provide a direct handle on the numerous non-gravitational phenomena that affect the physics of the intra-cluster medium.}
   {We present a proof-of-concept to model cluster number counts, that bypasses the  explicit use of scaling relations. We rather implement the effect of several astrophysical processes to describe the cluster properties. We then evaluate the performances of this modelling for the cosmological inference.}
   {We developed an accelerated machine learning baryonic field-emulator, based on an extension of the Lagrangian Deep Learning and trained on the CAMELS/IllustrisTNG simulations. We then created a pipeline that simulates cluster number counts in terms of XMM observable quantities. We finally compare the performances of our model, with that involving scaling relations,  for the purpose of cosmological inference based on simulations.}
   {Our model correctly reproduces the cluster population from the calibration simulations at the fiducial parameter values, and allows us to constrain feedback mechanisms. The cosmological-inference analyses indicate that our simulation-based model is less degenerate than the approach using scaling relations.}
   {This novel approach to model observed cluster number counts from simulations opens interesting perspectives for cluster cosmology. It has the potential to overcome the limitations of the standard approach, provided that the computing resolution and the volume of the simulations will allow a most realistic implementation of the complex phenomena driving cluster evolution.}

   \keywords{Cosmology: Galaxy clusters, X-ray astronomy, Cosmological simulations, Machine Learning.}

   \maketitle
%

\section{Introduction}

One of the most intriguing open question of modern cosmology, the nature of dark matter (DM), was raised almost a century ago by the study of velocities in galaxy clusters \citep{zwicky_rotverschiebung_1933} and, so far, remains unanswered. At the turn of the third millennium,  extensive supernova studies showed that, in the local Universe, expansion is accelerating \citep{riess_observational_1998, perlmutter_constraining_1999}. This indicates that the present-day Universe is under the influence of some ``negative pressure'', the so-called dark energy (DE). The matter-energy budget of the local Universe, thus only allows for a few percent of luminous baryonic matter.
\par To date, the simplest cosmological framework that accounts for the largest sample of observational facts is the $\Lambda$CDM model, within the framework of General Relativity, where the DE is the cosmological constant $\Lambda$. 
We have now entered the era of precision cosmology, where large and deep extragalactic surveys are multiplying: \textit{eROSITA} \citep{predehl_erosita_2021}, DESI \citep{desi_collaboration_overview_2022}, \textit{Euclid} \citep{collaboration_euclid_2024}, LSST \citep{ivezic_lsst_2019}, CMB-S4 \citep{abazajian_cmb-s4_2019}, SKA \citep{maartens_cosmology_2015} and \textit{ATHENA} \citep{nandra_hot_2013}. These facilities should allow testing theoretical models beyond $\Lambda$CDM, by combining various cosmological probes.

\par Galaxy clusters hold a privileged position in this quest. As the ultimate product of hierarchical structure formation, clusters, the nodes of the cosmic web, are sensitive to both the growth of structures and to the geometry of the Universe. In particular, they occupy the high end of the halo mass function (HMF), which is very sensitive to several cosmological parameters, especially $\Omega_m$ and $\sigma_8$.  This has motivated numerous cluster number count studies for more than three decades \citep{henry_extended_1992, vikhlinin_chandra_2009, mantz_observed_2010, bocquet_cluster_2019, garrel_xxl_2022}. The detection of the intra-cluster gas (ICM) in X-ray is much less subject to projection effects than galaxy densities in the optical; moreover,  ab initio modelling of the ICM is straightforward. For these reasons, X-ray cluster cosmology has been very productive. Future prospects for this waveband are exciting: from the  eROSITA all-sky survey \citep[$z\leq 1$; following the first survey analysis from][]{ghirardini_srgerosita_2024}, to \textit{ATHENA} in the next decade \citep[$1<z<2$;][]{cerardi_cosmological_2024}. Similarly, surveys in the millimeter wavelength, also targeting the ICM through the Sunyaev-Zel'dovich effect, offer a promising avenue, from SPT currently \citep{bocquet_multiprobe_2025} to SPT-3G \citep{sobrin_design_2022} and later CMB-S4.
\par The HMF predicts the number of clusters formed per unit mass and volume as a function of cosmic time. Because cluster masses are not directly measurable, scaling relations are used to relate X-ray luminosities or temperatures to mass, assuming hydrostatic equilibrium or using measurements from gravitational lensing to calibrate the mass scale. 
The slope and evolution of the scaling relations can be predicted from a purely self-similar gravitational collapse of matter overdensities \citep{kaiser_evolution_1986}. 
However, observed and simulated cluster populations present significant deviations from the expected scaling relations \citep[e.g. ][]{maughan_self-similar_2012, adami_xxl_2018, bulbul_x-ray_2019, bahar_erosita_2022}. The reason is twofold. First, each cluster is affected by its connections to the cosmic web, leading to non-spherical collapse and merger events, hence to departure from equilibrium \citep{arnaud_universal_2010, mantz_weighing_2016}. Second, strong radiative cooling occurs at the cluster centres and non-gravitational processes inject energy into the ICM  (from active galactic nuclei, or AGN, and supernova feedback, turbulence and magnetic fields). Fitting power-laws with intrinsic dispersion (possibly as a function of redshift) is an easy way to encapsulate all non-gravitational effects, in order to render the global properties of a cluster sample.  However, these empirical scaling-relation coefficients are  numerous (see table \ref{tab:compare_studies}) and degenerate with cosmology; e.g. are clusters in a given mass range not detected because they have not formed yet, or because they are under-luminous, or too extended?  Moreover, scatter (and its evolution) is a key parameter in the relations but is difficult to measure because of selection effects. All in all, while useful for implementing the necessary mass-observable connections in the cosmological analysis,  scaling relations are over-parametrised and do not allow direct insights into the individual non-gravitational  processes that galaxy clusters experience. 
\par To date, all cluster count analyses used empirical scaling relations in their approach to cosmology. \citep[e.g][ with  mass and luminosity, respectively]{vikhlinin_chandra_2009, mantz_observed_2010}. \cite{clerc_cosmological_2012, clerc_cosmological_2012-1} proposed to forward model the theoretical number counts from the HMF down to the clusters   properties, as registered by the detector (ASpiX method). In this new approach, the cluster population is summarised into a 3D X-ray Observable Diagram (XOD) combining count-rate, hardness ratio and redshift ($CR$, $HR$, $z$), analogous to a flux, colour, redshift diagram in X-rays. Although the subsequent cosmological analysis bypasses the calculation of the individual clusters masses and provides better control on the measurement errors, it still relies on the scaling relation formalism  for the likelihood computation. 
Only implicit inference methods ought to eliminate this component.
\par In a subsequent paper, we tested simulation-based inference techniques to perform number counts analysis \citep{kosiba_cosmological_2024}; we only relied on scaling relations for modelling the detected cluster population, but did not require any during inference. In the present work, our goal is to achieve the modelling of the physical cluster properties used in the cosmological inference, without the intermediate step involving scaling relations. 
\par  Hydrodynamical simulations offer a way to model clusters that takes into account the majority of the above-mentioned environmental and non-gravitational processes for a given set of cosmological parameters. They moreover implicitly carry information on the HMF and on the scaling relations. However, they are numerically too expensive to be used during the inference. To overcome this issue, we developed an accelerated simulation framework, powered by machine learning (ML) tools. Specifically, we combine GPU-accelerated DM-only (DMO) simulations with a fast baryonification technique, calibrated on hydrodynamical simulations. In a last step, we connect this model with the cosmological inference scheme from \cite{kosiba_cosmological_2024}. Of course, the final success of the proposed methodology very much depends on the degree of realism of the simulations, but in the present paper, we intend to remain at the proof-of-concept level. We list below  three important issues that constitute the basis of the paper.
\begin{enumerate}
    \item \textbf{Can we learn an accurate and fast ML baryonic field emulator to model cluster number counts?} We describe the ML approach adopted in section \ref{sec:fastsimulations}, along with the simulation pipeline that embeds it. The results are presented in section \ref{sec:results}.
    \item \textbf{What is the relevance of this approach for the inference?} We detail the inference stages in section \ref{sec:SBI}, and combine it with the simulation-based model to present its results in section \ref{sec:results}.
    \item \textbf{To which extent are simulations realistic?} This question is currently an obvious showstopper for the application of the method to real observations. We discuss the problem from several viewpoints in sections \ref{sec:results} and \ref{sec:discussion}.
\end{enumerate}
\section{Fast simulations with extended Lagrangian Deep Learning}\label{sec:fastsimulations}
In this section we describe a simulation-based forward model for cluster number counts. The model relies on DMO simulations and baryonification with Lagrangian Deep Learning (LDL), that makes it cheaper to run than full hydrodynamical simulations. We produce X-ray light cones from the emulated volumes and perform cluster detection and caracterisation.
\subsection{Hydrodynamical simulations}

\begin{table*}
\centering
\caption{Summary of the training data, and simulation parameters in the CAMELS dataset}\label{tab:camels_params}
\begin{tabular}{c|cc}
\hline
Simulation set & \makecell{CV50 (fiducial)} & \makecell{LH25 (conditioning)} \\ \hline \hline
\makecell{Number of simulations (training/validation)} & 27 (21/6) & 499 (400/99) \\ 
\makecell{Box volume ($h^{-3}$Mpc$^3$)} & 50$^3$ & 25$^3$ \\ 
\makecell{Redshift used in training} & 0.21 & 0.1, 0.15, 0.21, 0.27, 0.33, 0.40, 0.47, 0.54\\ 
\makecell{Training steps using the set} & base LDL & \makecell{extended LDL: conditioning and $z$ dependancy} \\ \hline
$\Omega_m$ &  $0.3$ & $\mathcal{U}(0.1, 0.5)$\\ 
$\sigma_8$ &  $0.8$ & $\mathcal{U}(0.6, 1.0)$ \\ 
$\ln(A_{SN1})$  & $\ln(1.0)$ & $\mathcal{U}(\ln(0.25), \ln(4.0))$ \\ 
$\ln(A_{SN2})$  & $\ln(1.0)$ & $\mathcal{U}(\ln(0.5), \ln(2.0))$ \\ 
$\ln(A_{AGN1})$ & $\ln(1.0)$ & $\mathcal{U}(\ln(0.25), \ln(4.0))$\\ 
$\ln(A_{AGN2})$ & $\ln(1.0)$ & $\mathcal{U}(\ln(0.5), \ln(2.0))$ \\ \hline
\end{tabular}
\tablefoot{CAMELS/IllustrisTNG sets used in this study. We describe in the upper part of the table the main characteristics and usage of each simulation set. We detail in the lower part the six parameters of interest for this work. For cosmology, the remaining parameters are set to the values from \cite{planck_collaboration_planck_2020}. Many other simulation parameters exist to monitor subgrid physics, but are kept fixed in this first version of CAMELS.}
\end{table*}

We use the CAMELS dataset \citep{villaescusa-navarro_camels_2023} to train our ML emulator. It contains thousands of hydrodynamical simulation boxes, run with different codes and for various cosmological and astrophysical parameters. As such, it is a well suited simulation set for ML projects. We here only consider the CAMELS/IllustrisTNG \citep{pillepich_simulating_2018,nelson_illustristng_2019} suite, specifically using the subsets:
\begin{itemize}
    \item Cosmic Variance (CV): 27 boxes of size 50 $h^{-1}$Mpc, all at the fiducial parameters values.
    \item Latin Hypercube (LH): 1000 boxes of size 25 $h^{-1}$Mpc, each one with a unique parameters set ($\Omega_m, \sigma_8, A_{AGN1}, A_{AGN2}, A_{SN1}, A_{SN2}$). The four latter parameters tune the astrophysical feedback, respectively acting on the AGN energy accumulation rate, the AGN burstiness, the SN energy injection rate and the SN wind speed. The fiducial value and range of variation of all the parameters are shown in table \ref{tab:camels_params}. We use a subset of 499 simulations from the full LH set.
\end{itemize}
Although the density of sampled points in the parameter space remains low, the LH strategy provides a quasi-uniform sampling, as shown in appendix \ref{app:LH_sampling}. While a larger number of simulations is generally preferable, we note that the LH set has been successfully used to develop conditioned emulators \citep[e.g. ][]{hassan_hiflow_2022}. In addition, the model we describe and train in \ref{sec:baseLDL} features a very shallow architecture, making it well-suited for use with a small training dataset. If one were to increase the number of free parameters, the Sobol sequence sampling might offer a solution to cover high dimensional spaces \citep[which has already been applied in the CAMELS subset with 28 parameters, see][]{ni_camels_2023}.
\par In this work, we focus on modelling the ICM electron number density $n_e$ and temperature $T$, which allows us to compute the X-ray emission of the gas (see section \ref{sec:sicc_pipeline}). We pre-process the input simulations to obtain these quantities on a regular grid, similarly to \citep{villaescusa-navarro_camels_2022}. We use the Cloud-in-Cell (CIC) algorithm to spread the simulated particles in comoving voxels. For both CV and LH, we use a voxel size of $0.39 h^{-1}$Mpc, which sets the working resolution throughout all this article.
\par In our simulation-based forward model, we evolve the DM with a fast Particle-Mesh (PM) approach, in order to accelerate further the field emulation. As a result, we resimulate the DM fields of all CAMELS/IllustrisTNG CV and LH with the code \textsc{JaxPM}\footnote{\url{https://github.com/DifferentiableUniverseInitiative/JaxPM}}, starting from the same initial condition at $z=6$. We train our ML emulator to mock the baryonic properties from the PM-approximated DM instead of the original DM field. The PM method induces a smoothing at small scales, and several techniques have been employed to compensate this effect \citep{dai_gradient_2018, lanzieri_hybrid_2022}. We do not use these methods here: our ML surrogate hence has the double objective of correcting the PM approximation and modelling accurately the baryonic properties. 
\par Lastly, it is important to note here that we train our model mainly on groups and lightweight clusters. Indeed, the small box size of CAMELS simulations prevents the formation of very massive halos. According to \cite{lee_zooming_2024}, only $\sim$15\% of LH boxes form a cluster with mass $> 10^{14.5} h^{-1}$M$_{\odot}$.

\subsection{Fast baryonification}

\subsubsection{Base Lagrangian Deep Learning}\label{sec:baseLDL}
We use the Lagrangian Deep Learning framework to quickly emulate $n_e$ and $T$ from DM fields \citep[LDL, ][]{dai_learning_2021}. This approach uses two kinds of learnable layers: one of particle displacement, and one of non-linear activation. The first one acts on the DM simulated particles, moving them following a modified potential. We use two consecutive displacement layers in our LDL model. The activation layer, unique in our model, introduces the non-linearities of baryonic processes. Compared to deep generative models (such as U-nets) LDL is a very lightweight approach, and follows,by design, physical principles (rotation and translation invariance, particles moving along a potential).
\par Formally, if we consider an input overdensity field $\delta(\mathbf{x})$ (which is the DM overdensity only if this is the first layer), we first model a source term:
\begin{equation}
    f(\mathbf{x}) = (1+\delta(\mathbf{x}))^\gamma,
\end{equation}
with $\gamma$ a learnable parameter.
We use a highly flexible radial Fourier filter based on a B-Spline $\mathscr{S}(\Xi, k)$, inspired from \cite{lanzieri_hybrid_2022}:
\begin{equation}
  \mathbf{\hat{O}_{\mathscr{S}}}(k) = 1+\mathscr{S}(\Xi, k),
\end{equation}
where $\Xi$ are the B-Spline paramters, learnable. The displacement field, applied to the input particles, then writes:
\begin{equation}
    \diff \mathbf{x} = \alpha \nabla \mathcal{F}^{-1}
    \left(\mathbf{\hat{O}_{\mathscr{S}}}(k) \hat{f}(\mathbf{k}) \right),
\end{equation}
with $\alpha$ an additional learnable parameters, and $\mathcal{F}^{-1}$ the inverse Fourier transform. After repeating two times this displacement procedure, the activation is performed with a Rectified Linear Unit (ReLU):
\begin{equation}
    F(\mathbf{x'}) = 
    \text{ReLU}\left(b_1(1+\delta(\mathbf{x'}))^\mu - b_0\right),
\end{equation}
with $b_0, b_1, \mu$ completing the list of learnable parameters, which we all regroup under the symbol $\Theta$. Note that the output of the LDL is not a set of particles but a field on a comoving grid.

\par The LDL layer parameters, similar to neural network weights, are trained using the target property from a hydrodynamical simulation. As the end goal of our baryonification technique is to reproduce the X-ray emissivity ($\propto n_e^2 T^{1/2}$) of the gas in clusters, the model is trained using the following loss functions:
\begin{align}
    \mathcal{L}_{n_e} = \sum_i \left\lVert \mathbf{O_s} *
    \left[ \left( n^2_{e_{LDL}}(x_i)
    - n^2_{e_{true}}(x_i)\right) T^{1/2}_{true}(x_i)\right]
    \right\rVert \rho_{DM}(x_i), \\
    \mathcal{L}_{T}  = \sum_i \left\lVert \mathbf{O_s} *
    \left[  n^2_{e_{true}}(x_i) \left( T^{1/2}_{LDL}(x_i)
    - T^{1/2}_{true}(x_i)\right) \right]
    \right\rVert \rho_{DM}(x_i),
\end{align}
which use a smoothing operator $\mathbf{\hat{O}_s}(k) = 1 + k^{-n}$. The value for $n$ follows the prescription from \cite{dai_learning_2021}.
We here use the CAMELS/IllustrisTNG CV set, at $z=0.21$, a choice motivated by the usual peak of the redshift distribution of clusters in current X-ray surveys.

\subsubsection{Extended LDL}

The LDL approach we have sketched is trained to reproduce a specific simulation model, i.e. the fiducial model at a specific redshift. However, during the cosmological inference, we want to vary the cosmological parameters, as well as the feedback parameters that may impact the intergalactic gas, which are $A_{AGN1}, A_{AGN2}, A_{SN1}, A_{SN2}$ in CAMELS/IllustrisTNG. We hence need an extended LDL model, whose weights $\Theta$ are conditioned on the simulation parameters $\theta_{sim}$=($\Omega_m, \sigma_8, A_{AGN1}, A_{AGN2}, A_{SN1}, A_{SN2}$). Then, as we need to populate lightcones, we also want this baryonification to depend on the redshift.
\par
\begin{figure*}
  \centering
  \includegraphics[width=.8\textwidth]{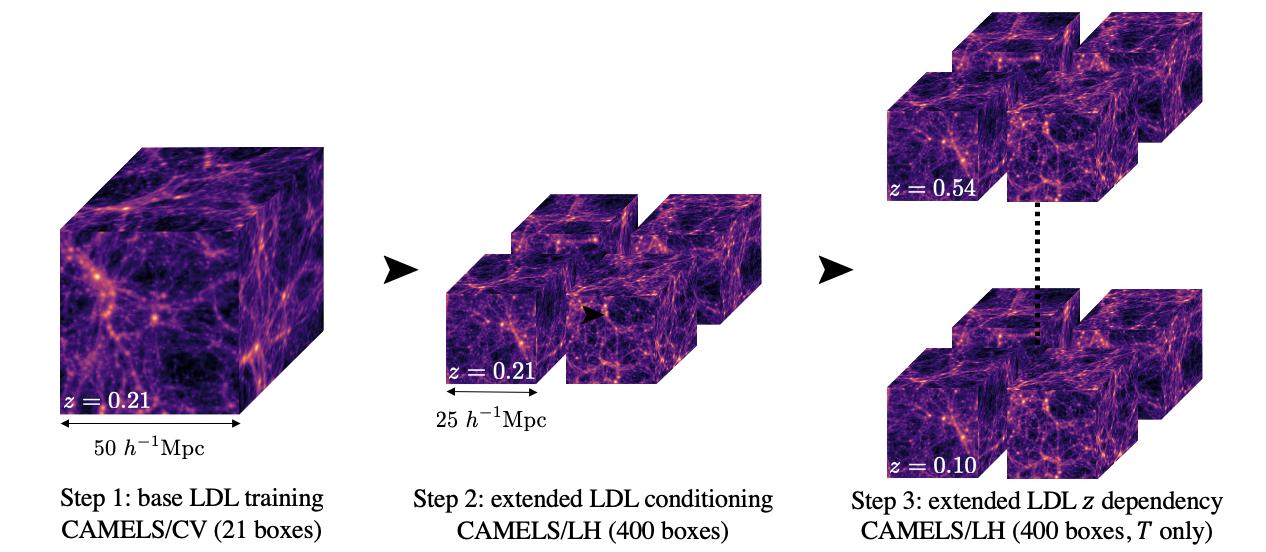}
    \caption{Training stages of the extended LDL. We first train the base LDL parameters on the large volumes available for the fiducial model (CAMELS/CV), at $z=0.21$. We then condition the LDL parameters on the cosmological and astrophysical parameters, using the numerous boxes from the CAMELS/LH set. While the $n_e$ model performs equally at all redshifts, the $T$ emulator has to be retrained separately at all available redshifts.}
    \label{fig:training_stages_eLDL}
\end{figure*}

\begin{figure}
  \centering
  \includegraphics[width=.48\textwidth]{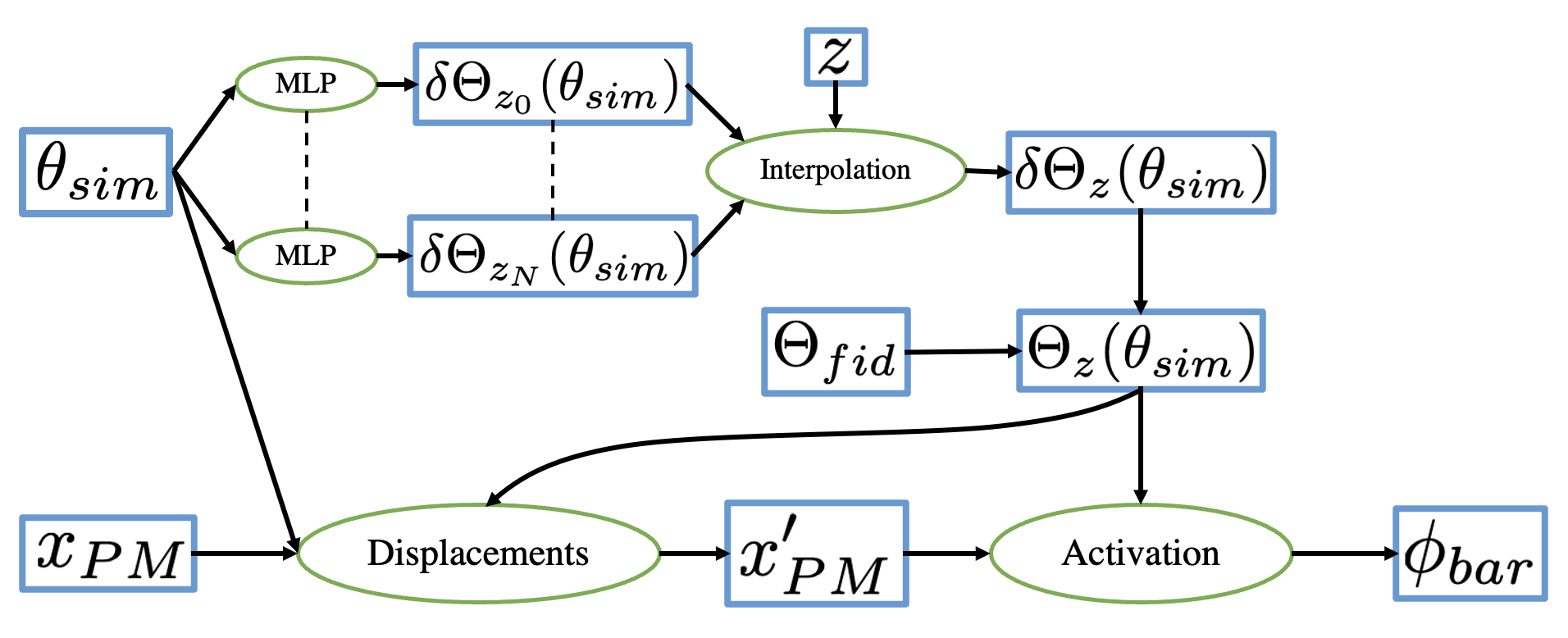}
    \caption{Scheme of the extended LDL. Blue rectangles denote quantities and green ovales denote transformations. The bottom raw is the base LDL from \cite{dai_learning_2021}. Our extensions allow the baryon pasting to be conditioned on simulation parameters as well as on the redshift.}
    \label{fig:scheme_eLDL}
\end{figure}

 We retain the base LDL parameters trained on CV, that we coin $\Theta_{fid}$. We use a simple multi-layer perceptron (MLP) to output a weight variation $\delta\Theta$. Following a meta-learning approach, we now train the MLP on the LH set, passing to the LDL model the weights:
\begin{equation}
    \Theta = \Theta_{fid} + \delta\Theta(\theta_{sim}).
\end{equation}
This second step of training is also restricted to $z=0.21$. To make the baryonification model dependent on $z$, we duplicate the MLP $\delta\Theta(\theta_{sim})$ and retrain it separately at the available redshifts in the LH set. For a given $z$, surrounded by $z_k$ and $z_{k+1}$, the closest redshifts for which a model is trained, we compute $\delta\Theta_z(\theta_{sim})$ by linear interpolation between $\delta\Theta_{z_k}(\theta_{sim})$ and $\delta\Theta_{z_{k+1}}(\theta_{sim})$. We find that by emulating $n_e$ in comoving coordinates, the accuracy of our model is not improved by adding the dependence on $z$ ($z<0.5$, see Sec. \ref{sec:sicc_pipeline}). Consequently, we bypass the interpolation step for $n_e$. To the contrary, the $T$ fields show a strong dependency on $z$, which can be explained by the strong influence of feedback on the temperature, so the redshift-conditioning is required for $T$. We resume the successive training stages in figure \ref{fig:training_stages_eLDL}. The functional scheme of the extended LDL is sketched in the figure \ref{fig:scheme_eLDL}.

\subsection{Simulation-based pipeline}\label{sec:sicc_pipeline}

\begin{figure*}
  \centering
  \includegraphics[width=\textwidth]{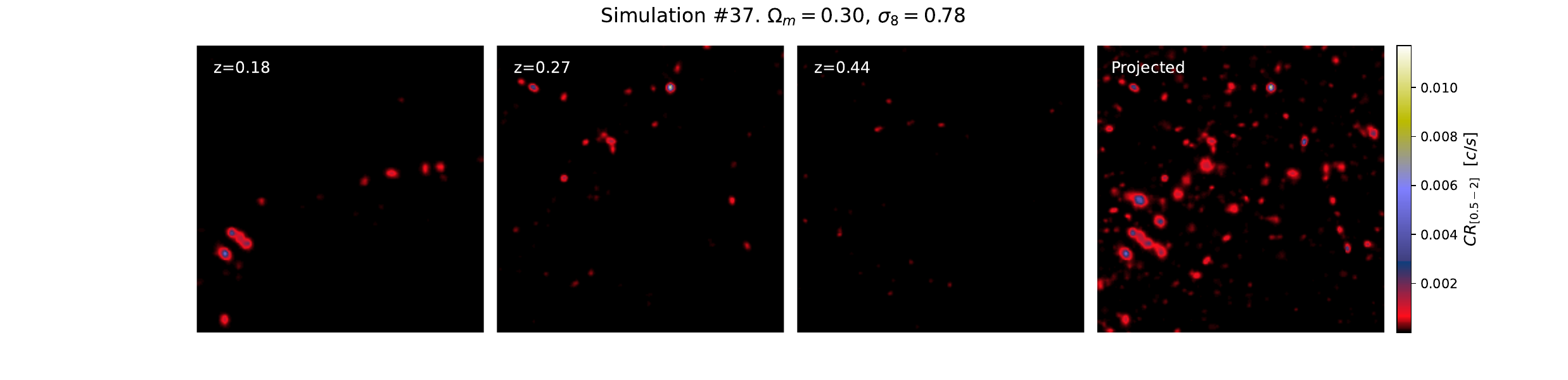}
  \includegraphics[width=\textwidth]{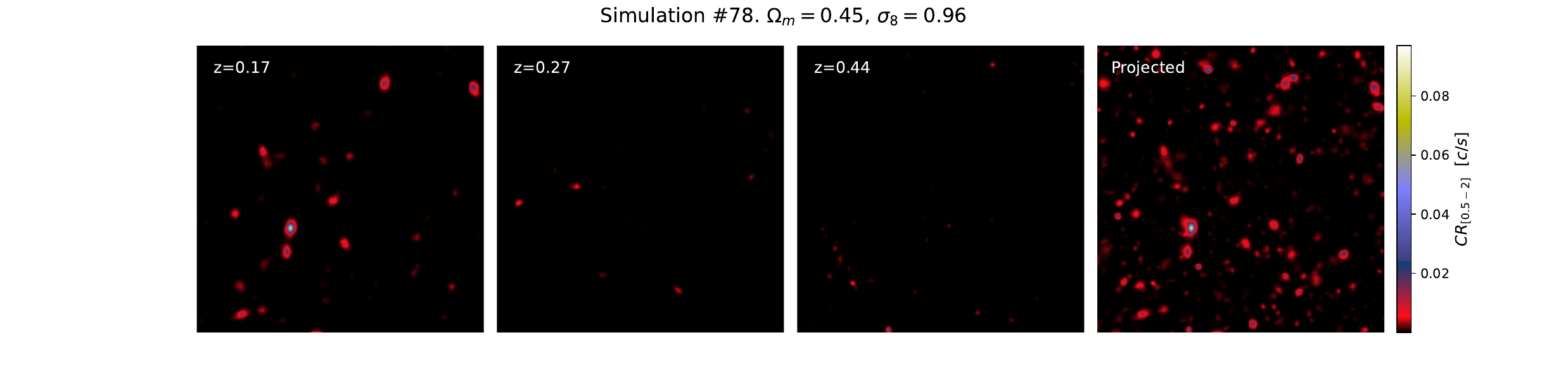}
    \caption{X-ray mock examples generated with the extended LDL. Each row consist of the same light cone, with three boxes individually projected (left and middle), and the full projected lightcone with its 12 simulations boxes (right). The $CR$ corresponds to a XMM-\textit{Newton} EPIC/mos1+mos2+pn count-rate (without Poisson noise), in the band [0.5-2] keV.}
    \label{fig:SICC_mocks}
\end{figure*}

\begin{figure*}
  \centering
  \includegraphics[width=\textwidth]{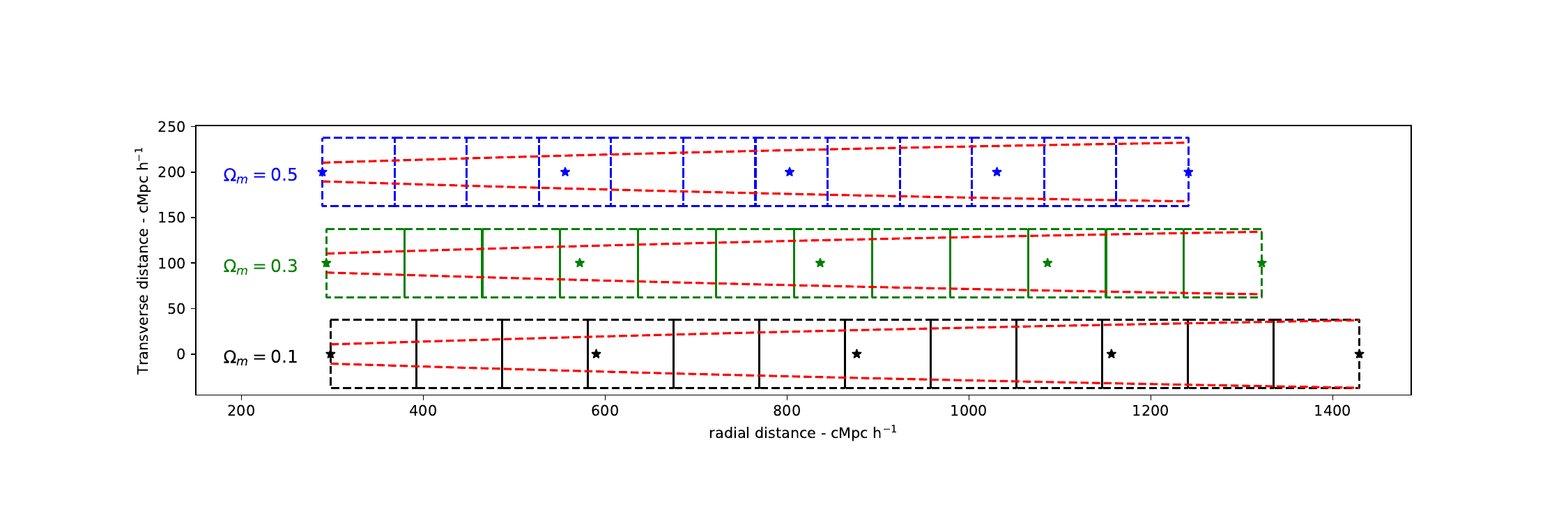}
    \caption{Lightcone profiles for $\Omega_m=0.1,0.3,0.5$ (resp. black, green, blue), intersected with the 20 deg$^2$ observed region (dashed red cone). The 12 adjacent periodic boxes (dashed rectangles) in each lightcone share the same comoving dimensions. Also shown with stars are the redshifts $(0.1, 0.2, 0.3, 0.4, 0.5)$ for each cosmology.}
    \label{fig:lightcone_geo}
\end{figure*}

We use the extended LDL to produce mock X-ray observations. We generate 20 deg$^2$ lightcones extending from $z=0.1$ out to $z=0.5$. These boundaries are due to a current limitation of our code, which does not support yet parallelization of one simulation over several GPUs. As a result, we have tuned the redshift range so that one simulation almost fill up the memory of a 32Go V100 GPU. One lightcone is made of 12 adjacent periodic boxes of equal comoving size, each one evolved with different random initial conditions at $z=99$. The depth of the boxes varies with $\Omega_m$, and so the number of particles and voxels. Several lightcones profiles are displayed in figure \ref{fig:lightcone_geo}. We evolve DMO fields with \textsc{JaxPM}, and then paint in voxels the electron number density and the temperature with the extended LDL, all at a fixed resolution of 0.39$h^{-1}$Mpc.
\par We then compute the X-ray emission of each voxel. We assume the intracluster gas is at a fixed metallicity of $Z=0.3Z_\odot$, and we use \texttt{pyatomDB}\footnote{\url{https://atomdb.readthedocs.io/}}\citep{foster_updated_2012} to compute the bremsstrahlung emission spectrum. We have developed \texttt{xrayflux}\footnote{\url{https://github.com/nicolas-cerardi/xrayflux/}}, a convenient wrapper of \texttt{pyatomDB} that redshifts the spectra and convolves them by  the instrument response. We use the XMM-\textit{Newton}/EPIC response files, and we obtain count rates (CR) in the bands [0.5-1] and [1-2] keV (observer frame), similarly to the framework used in the XXL Survey \citep{pierre_xxl_2016}. We did not model the galactic absorption, which is negligible when observing far enough from the galactic plane, like in XXL. For wider surveys, the modelling of the galactic absorption and of its variations across the covered area would be necessary \citep[as for instance in][]{clerc_srgerosita_2024}. For simplicity, we do not include X-ray AGNs in the field, and do not add Poisson noise on the photon counts. We project separately the boxes along the line of sight, so that we obtain 12 X-ray maps in each energy band, with a resolution of $\sim 1$'.
\par The detection is performed independantly on each map along the line of sight, over the total [0.5-1]+[1-2] energy band, using \texttt{sep} \citep{barbary_sep_2016}, a python implementation of \texttt{SExtractor} \citep{bertin_sextractor_1996}. For each detected object, we sum the $CR_{[0.5-2]}$ in its corresponding segmentation mask to obtain the integrated source $CR$. We retain all sources with $CR_{[0.5-2]}>0.02$ c/s, a flux limit that gives a density of cluster similar to the XXL C1 sample with the traditional forward model \citep{kosiba_cosmological_2024}. For the selected sources, we compute their hardness ratios $HR\equiv CR_{[1-2]}/CR_{[0.5-1]}$, a tracer of the cluster temperature and redshift. The adopted detection method is simple and could be upgraded to better reproduce real-observation processing \citep{pacaud_xmm_2006}, but is sufficient for the purpose of this proof-of-concept paper. Moreover, we mention that projection effects within a lightcone are ignored in the detection process. We have omitted the Poisson noise and neglected measurement errors on the $CR$ and the $HR$, which has an impact on the number of objects detected. Some X-ray mocks are shown as example in figure \ref{fig:SICC_mocks}. We finally regroup the objects detected in 10 independent lightcones to populate a 3D XOD, representative of a 200 deg$^2$ and 10ks depth survey carried out with XMM-\textit{Newton}. These XOD serve as the input statistic used during the inference.
\par This pipeline benefits from GPU acceleration until the detection step. Thanks to the PM simulations and the eLDL, the whole generation process is very fast: $\sim$100 seconds are required to produce an XOD of a 200 deg$^2$ survey, evolving a total of $\sim10^9$ DM particles.

\section{Simulation-based cosmological inference}\label{sec:SBI}
We described in the previous section a fast forward model based on cosmological simulations. From this, analytical scaling relations are required to write the likelihood and to perform a traditional inference scheme like MCMC. However, here, our simulations do not provide us with such scaling relations, and prevents us from using an explicit likelihood. We are therefore turning to  simulation-based inference (SBI), a new field that emerges thanks to the advances in ML. Specifically, the use of neural density estimators \citep[such as][]{bishop_mixture_1994, rezende_variational_2016} enables us to learn the probability density of interest. We here detail our approach, inspired from \cite{kosiba_cosmological_2024}.
\subsection{Neural compression}

As a first step, we perform neural compression on our XODs. The motivation for this is that SBI techniques usually work better on low dimension statistics. We build a convolutional neural network with ResNet blocks \citep{he_deep_2015}, which takes the XODs $x$ as input (with the redshift dimension as image channels) and outputs the neural compression $y$, a vector containing six scalars, the size of the simulation parameter vector $\theta_{sim}$. We use a rather shallow architecture, with 4 ResNet blocks (3 convolution layers each) followed by one MLP (3 dense layers). We train our CNN to retrieve the latter using a mean-squared error loss, a choice that can lead to biased posterior \citep{lanzieri_optimal_2024, jeffrey_likelihood-free_2020}. As we do not apply the posterior estimation on real data, and are primarly interested by the size of the constraints, this choice is not critical in this paper.
\par We generate with our pipeline a large sample of 48000 XODs, each one with a random set of simulation parameters. 32000 XODs are used for the training, 4000 for the validation, 4000 for testing. The remaining 8000 are used for the inference. We apply a normalization on both the input XODs and the target parameters such that all fields seen by the ResNet are between 0 and 1. The generation of these datasets necessitates $\sim 2000$hGPU, while the ResNet training and the inference only requires $\sim 1$hGPU.

\subsection{Neural Posterior estimation}
We then directly learn the posterior from the compressed statistics $y$, using the neural posterior estimation (NPE) technique \citep{papamakarios_fast_2018}. Unlike neural likelihood estimation \citep[NLE,][]{wood_statistical_2010}, the NPE is amortized, meaning that we do not need to retrain our density estimator when we have a new observation $y_0$. One downside of the NPE is that the prior is encoded in the learnt density, and hence cannot be easily changed. We use a mixture density network (MDN) to emulate the conditional posterior $q_\varphi(\theta\mid y)$, with $\varphi$ the MDN weights. We use the package \texttt{sbi} \citep{tejero-cantero_sbi_2020} to train a MDN instantiated with 10 Gaussian components. For any $y_0$ unseen during training, we can directly sample the neural posterior $q_\varphi(\theta\mid y=y_0)$, show the high density regions and compute the uncertainties on each parameter. 

\section{Results}\label{sec:results}
In the first place, we introduce the analytical forward model to be compared with the simulation-based pipeline developped in this work. In a second step, we study the fidelity of the LDL surrogate simulations, relatively to the original hydrodynamical CAMELS. Afterwards, we look at the possibility to obtain posteriors using this accelerated simulation-based forward model.

\subsection{Analytical modelling}\label{sec:analyticalmodel}
As a baseline for comparison, we consider a traditional forward model that uses empirical scaling relations. In this paper, we are mainly interested in comparing the two methods on the inference task, but shall also compare the observed $\diff N/\diff z$ in order to quantify the general level of realism of our model. We keep the fiducial values of $\Omega_m$ and $\sigma_8$ from table \ref{tab:camels_params}, and retain the \cite{planck_collaboration_planck_2020} values for the remaining cosmological parameters. We then express the number counts in the mass-redshift plane with the HMF from \cite{tinker_toward_2008}. We use the scaling relations from \cite{pacaud_xxl_2018} to obtain the cluster temperature and luminosity, following which we compute the measured X-ray flux with the XMM-\textit{Newton}/EPIC instruments using \texttt{XSPEC} \citep{arnaud_xspec_1996}. As in the previous section, we only keep the objects with a $CR$ larger than $CR_{[0.5-2],lim}=0.02$ c/s. This does not ensure the selection functions to be strictly equal between both models, as the detection algorithm used on the simulations may induce some artefacts. We finally obtain the source density  in the ($CR$, $HR$, $z$) space, which we multiply by the survey solid angle, 200 deg$^2$. We compare the simulation-based and analytical forward models in table \ref{tab:compare_studies}.
\begin{table*}[]
\centering
\caption{Comparison of different number counts modelisation, in this work and from the literature.}
\begin{tabular}{cccc}
Study  & $z$ range & Observables & \makecell{Number of free parameters\\ (cosmo+nuisance)} \\ \hline \hline 
This work, simulation-based model  &  $0.1<z<0.5$  &   $CR,HR,z$           &   2+4      \\
This work, analytical model & $0.1<z<0.5$ &   $CR,HR,z$        &     2+6     \\ \hline
ROSAT \citep{mantz_observed_2010}    & $z<0.5$ &  $L,T,M_{gas},z$   &  4+27       \\
XMM-XXL \citep{garrel_xxl_2022}  & $0.05<z<1$  &  $CR,HR,z$           & 5+7     \\ 
eRASS1 \citep{ghirardini_srgerosita_2024} &  $0.1<z<0.8$ &  $CR, M_{WL}, \lambda, z$   &  4+22  \\  \hline
\end{tabular}
\tablefoot{The top part of the table shows the characteristics of the models used in this article. The bottom part quotes the main characteristics of models applied on real X-ray surveys.}\label{tab:compare_studies}
\end{table*}

\subsection{Extended LDL results}\label{sec:results_eLDL}
\begin{figure}
  \centering
  \includegraphics[width=8cm]{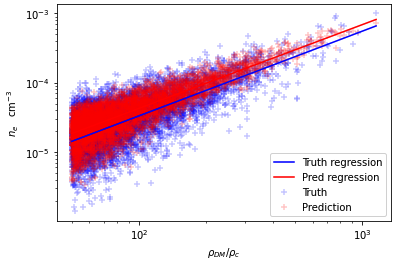}
        \caption{Relation between the electron number density $n_e$ and the DM density for the hydrodynamical simulations (blue crosses) and the LDL surrogate (red crosses), for the fiducial parameters and at $z=0.21$. The blue and red lines denote the mean locus (respectively for the hydrodynamical and the LDL-predicted voxels). The voxel deviation from the mean for both methods is quite correlated ($\rho_{corr}= 0.56$), but the LDL predictions do not reproduce the same scatter as in the original simulations.}
    \label{fig:pixelscatter}
\end{figure}

Although we are interested in clusters, we first take a look at the raw LDL results, i.e. at the voxel level. In figure \ref{fig:pixelscatter}, we show the voxel electron number density as a function of the DM density to critical density ratio, for the test simulations of the CV set, both for the original hydrodynamical boxes and their LDL-predicted counterparts. The dispersion of the LDL prediction, at the voxel level, is $(1.90 \pm 0.08) \times 10^{-5}$ cm$^{-3}$, significantly less than that of the original values, $(3.0 \pm 0.1) \times 10^{-5}$ cm$^{-3}$ (over 5$\sigma$ statistical difference). This means that our model does not fully capture the diversity of $n_e$ at a given $\rho_{DM}/\rho_c$. We moreover fit and display power-laws for both the target and predicted voxels. We compute the deviation of each voxel from the power law, and the correlation between the hydrodynamical and LDL deviations. The correlation coefficient is found to be $0.56$, which indicates that the LDL voxels are not simply randomly dispersed around the mean power-law. However, these results concern the properties of voxels only and do not yield much information on our primary focus, clusters, so we prevent ourselves to overinterpret the raw LDL predictions.

\par
\begin{figure*}
  \centering
  \includegraphics[width=6cm, trim=0 23.3cm 23.3cm 0, clip=true]{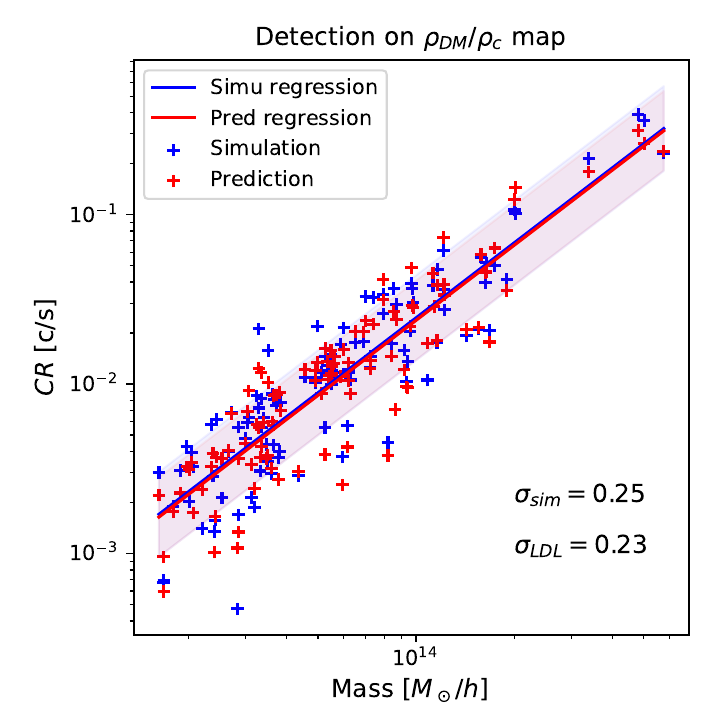}
  \includegraphics[width=6cm, trim=0 11.7cm 23.3cm 12.5cm, clip=true]{figures/haloes_SL_CR_pos_2.pdf}
  \includegraphics[width=6cm, trim=0 0.1cm 23.3cm 23.7cm, clip=true]{figures/haloes_SL_CR_pos_2.pdf}
    \caption{$CR-M$ scaling relation at $z=0.21$ in the CAMELS/IllustrisTNG simulations and in their LDL surrogate, for the fiducial model. These plots are made with the CV test set, 6 boxes of volume (50 $h^{-1}$Mpc)$^3$ each.\newline Left: direct comparison of the hydrodynamical simulated clusters (blue crosses) and their LDL counterpart (red crosses). The plain lines and shaded region indicate respectively the mean relation and scatter for each method. The scaling relation is well reproduced by the LDL method. 
    \newline Middle: Deviation of the hydrodynamical values and LDL prediction to the mean relation (blue points), and the 1:1 slope (dashed line). The strong correlation indicates that the LDL recovers more than just a scaling relation: it can learn why a specific cluster is over or under luminous, given its mass, thanks to its 3D DM distribution.
    \newline Right: histogram of the $CR$ prediction errors. At the cluster level, the LDL prediction appears unbiased, and with a scatter of 0.12 dex, smaller than the natural scatter of the $CR-M$ relation.}
    \label{fig:SICC_SL}
\end{figure*}

 Secondly, we look at the results on clusters, using the procedure described in \ref{sec:sicc_pipeline}. Although we do not use explicit scaling relations in our modelling, in principle the cluster population in the CAMELS simulations should emulate implicit scaling relations, a physical consequence of the collapse of DM overdensities. Once again, relying on the test simulations from the CV set at $z=0.21$, we compute the mock X-ray maps from both the original and the LDL simulations. We here run the detection on the $\rho_{DM}/\rho_c$ maps, to avoid any bias toward the true or predicted structures in the $CR$ maps. We compute the mass, the true $CR$ and the predicted $CR$ by integrating over the detection mask of each source. The measured mass here does not exactly relate to the spherical overdensity mass as in classical cluster studies. But as the detection uses a threshold of $\rho_{DM}/\rho_c=200$, it can be considered as an approximation of $M_{200c}$, without the spherical assumption and including potential contaminants on the line of sight. Still, we can compare together our hydrodynamical and LDL clusters since their mass estimates are consistent. In figure \ref{fig:SICC_SL}, we first show the $CR-M$ relation for each  cluster set. Our LDL-emulated cluster sample reproduces very well the normalization, slope and scatter of the clusters in CAMELS/CV. But our approach does more than just recovering the scaling relation: we again compute the deviation of each cluster  $CR$ to the mean relation, and find that the LDL and true deviations are very well correlated ($\rho_{corr}=0.88$). This shows that, thanks to the 3D modelling approach, our trained LDL is able to integrate information from the cluster environments (but not all), and how this affects cluster properties. In that sense, our modelling is superior to an explicit and empirical scaling relation. We also compute the error on the prediction (see figure, right panel), and find it to be two times smaller than the natural scatter of the $CR-M$ relation. However, this study cannot be reproduced for the LH set, given that for one specific set of simulation parameters, we only have one (25$h^{-1}$Mpc)$^3$ box, which does not provide sufficient statistics on clusters. This shows a limitation of the version of CAMELS we use here.

\par
\begin{figure*}
  \centering
  \includegraphics[width=12cm, trim=0 0 0 0, clip=true]{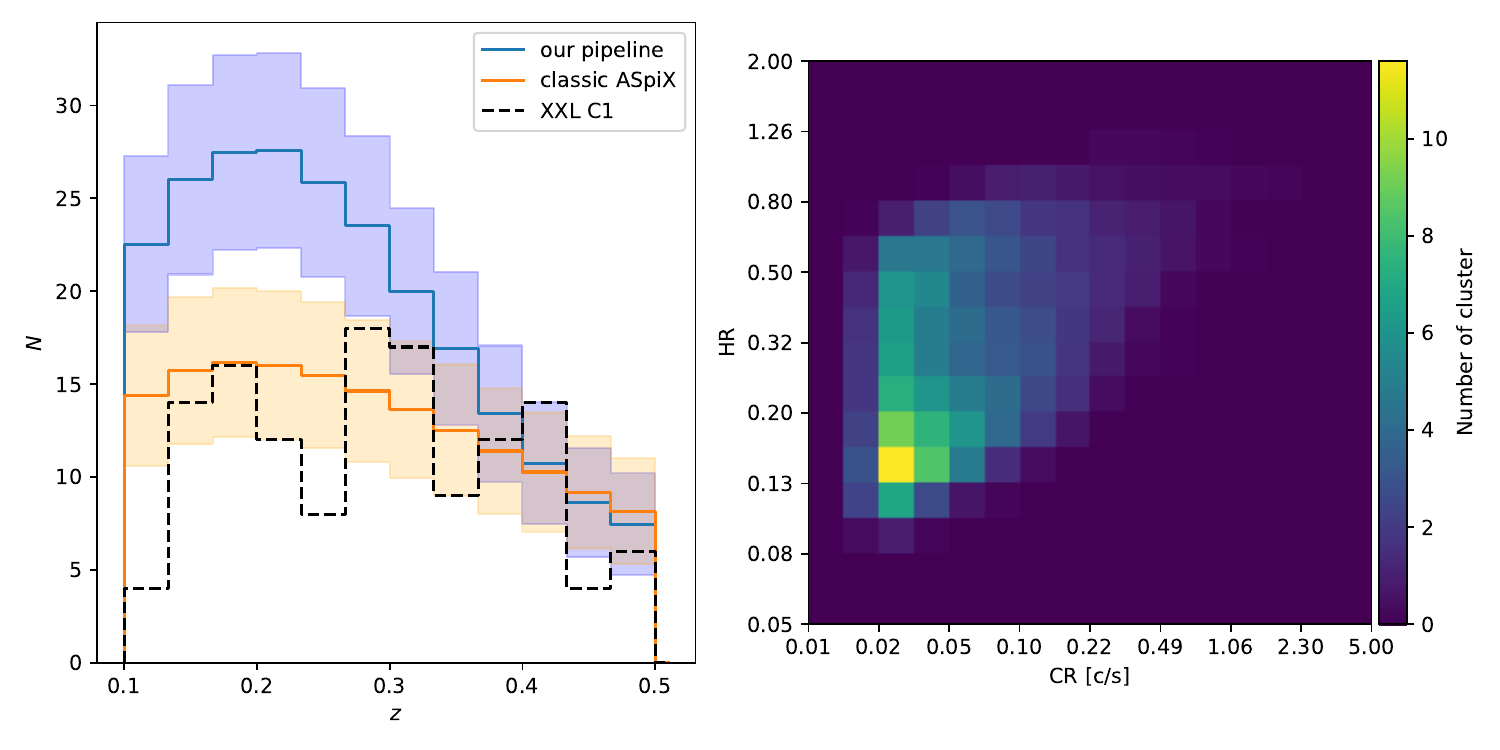}
    \caption{Left: Redshift distribution of cluster counts for the traditional forward model (with explicit scaling relations, orange histogram), and for the simulation-based model (with the LDL, blue histogram), here for a 50 deg$^2$ survey extending from $z=0.1$ to $z=0.5$. The shaded regions display the $3\sigma$ Poisson standard deviation for each model. Also shown is the observed distribution of C1 clusters in the XXL Survey (dashed black histogram), however one should keep in mind that the selection function is in this case more complex than a simple flux cut. We stress that this figure is not a goodness of fit: both models are calibrated independently. \\ Right: cluster population in the $CR$-$HR$ space for the simulation-based model, also for a 50 deg$^2$ survey extending from $z=0.1$ to $z=0.5$. The $CR$ and $HR$ in the simulation-based and scaling relations-based models are computed in different manners; hence we refrain ourselves to compare their distirbutions.}
    \label{fig:SICC_dndz}
\end{figure*}

 We now compare the cluster number counts from simulation and analytical models, in figure \ref{fig:SICC_dndz}. For our simulation-based model, we here generate 48 surveys covering 50 deg$^2$, with $z\in [0.1, 0.5]$, and measure the redshift distribution in each of them. We obtain theoretical number counts with the analytical model presented in section \ref{sec:analyticalmodel}, for a survey with the same dimensions. We also display the observed distribution from the XXL Survey, although one should keep in mind that it is not directly comparable, because the XXL selection function is more complex than the simple CR cut we have been using. While the simulated and analytical distributions share the same peak position (at $z=0.2$) and compatible number counts for $z>0.3$, we observe a significant discrepancy at lower redshifts, where our simulated-based number count is larger than the explicit modelling: the total number of clusters with our LDL simulations is 40\% higher than with the analytical model. This is surprising as we have shown previously that our LDL model well reproduces the fiducial cluster population from CAMELS. But we have left aside several important questions. Firstly, it is possible that the X-ray luminosity of simulated clusters is not realistic enough. \cite{pop_sunyaev-zeldovich_2022} showed that the X-ray luminosity in IllustrisTNG clusters is slightly higher than that of observed samples, in particular for the core-excised luminosity. We also note from this study that the $L_X - M$ relation of IllustrisTNG clusters appears to be low-scattered. Through selection effects, small differences in the simulated luminosities can cause notable changes in the $\diff N / \diff z$. In addition, we have no guarantee that the HMF arising from the CAMELS simulations is compatible with the functional fit from \citep{tinker_toward_2008}. Notably, it is known that baryonic processes affect the HMF \citep{bocquet_halo_2016, kugel_flamingo_2024}. Aside from this considerations, we recall that we have implemented very simple detection and property measurement processes, which are also likely to impact the number counts in output. In this work, we focus on the LDL acceleration and SBI for cluster cosmology, so we leave open these questions for now. Certainly, it will be necessary to consider them before applying our method on real data.

\subsection{Posteriors}
\begin{figure*}
  \centering
  \includegraphics[width=15cm]{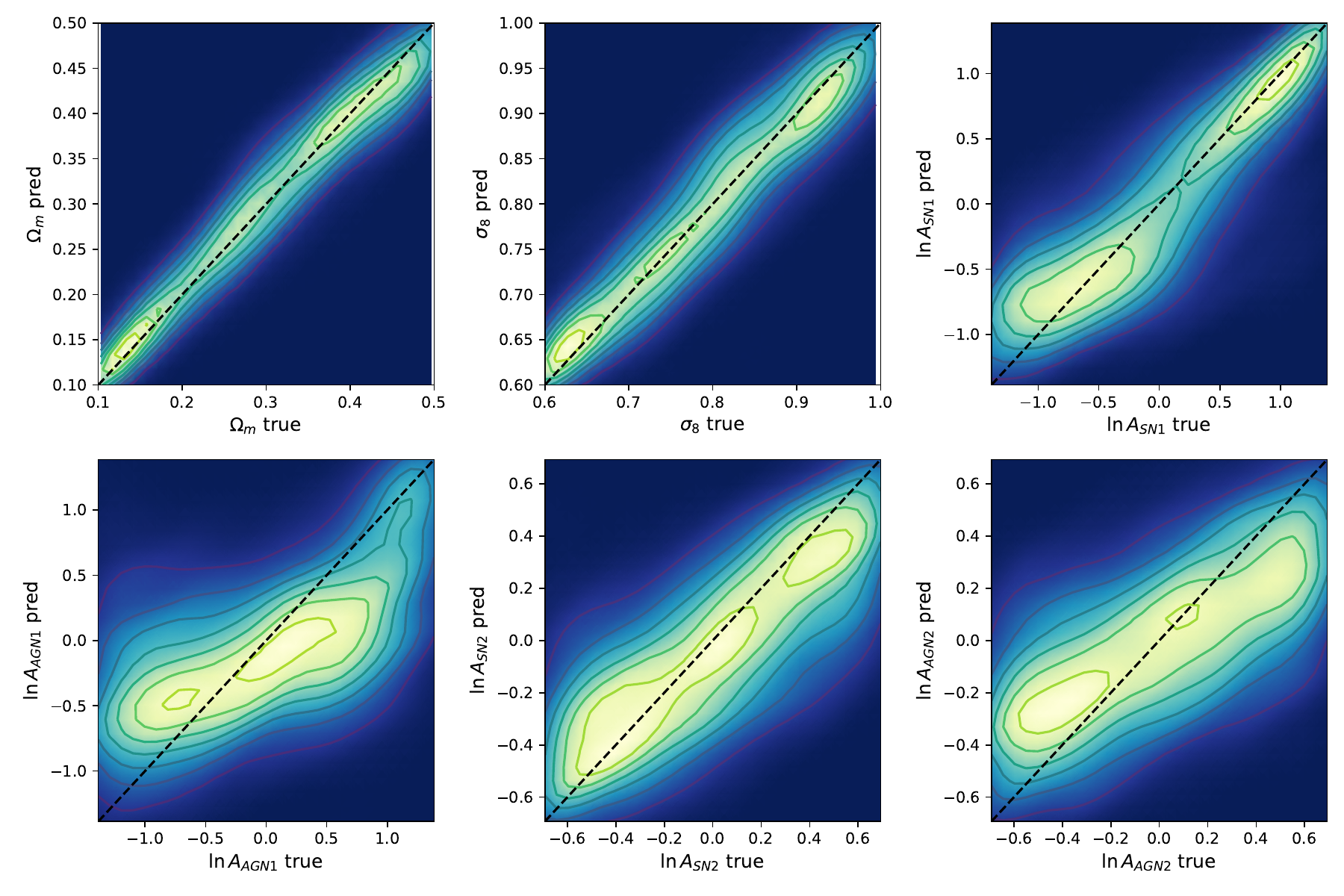}
    \caption{Accuracy of the ResNet regression for inferring $\Omega_m$, $\sigma_8$, $A_{SN1}$, $A_{AGN1}$, $A_{SN2}$, $A_{AGN2}$ from simulation-based XODs. The color and contours show the shape of the density of points (arbitrary colour scale, all densities are normalized). The dashed line show the 1:1 line (no error). We observe that the two cosmological parameters are well recovered. The SN feedback parameters are also rather well estimated, although with a stronger scatter. However, the ResNet struggles to retrieve the AGN feedback parameters, meaning that these parameters do not influence much our XOD statistics.}
    \label{fig:regression_RD}
\end{figure*}

We now present the results of the simulation-based inference with the ResNet compressor and the NPE method. We recall that in the SBI steps, our models are trained and tested on the large dataset of XODs (48000 cluster catalogues sampled) generated with the extended LDL and pipeline described in section \ref{sec:fastsimulations}. We first look at the regression performance of the ResNet model, trained and tested on our simulated XODs dataset varying the six simulation parameters considered in this work, in figure \ref{fig:regression_RD}. The ResNet retrieves well $\Omega_m$ and $\sigma_8$, as the points are distributed close to the 1:1  slope. The SN feedback parameters are also recovered, but with a larger scatter. However, the ResNet predictions are highly scattered for the AGN feedback parameters as currently implemented in CAMELS, and present biases near the edge of the sampled prior (see figure \ref{fig:regression_RD}, bottom left and bottom right panel). This indicates that our forward model is not very sensitive to these parameters, at least under the XOD statistics, and for a 200 deg$^2$ survey. The weak response of our LDL model to the AGN feedback variations in fact stems from the CAMELS dataset itself. We discuss this point in section \ref{sec:discussion_realdata}. 

\par
\begin{figure*}
  \centering
  \includegraphics[width=14cm]{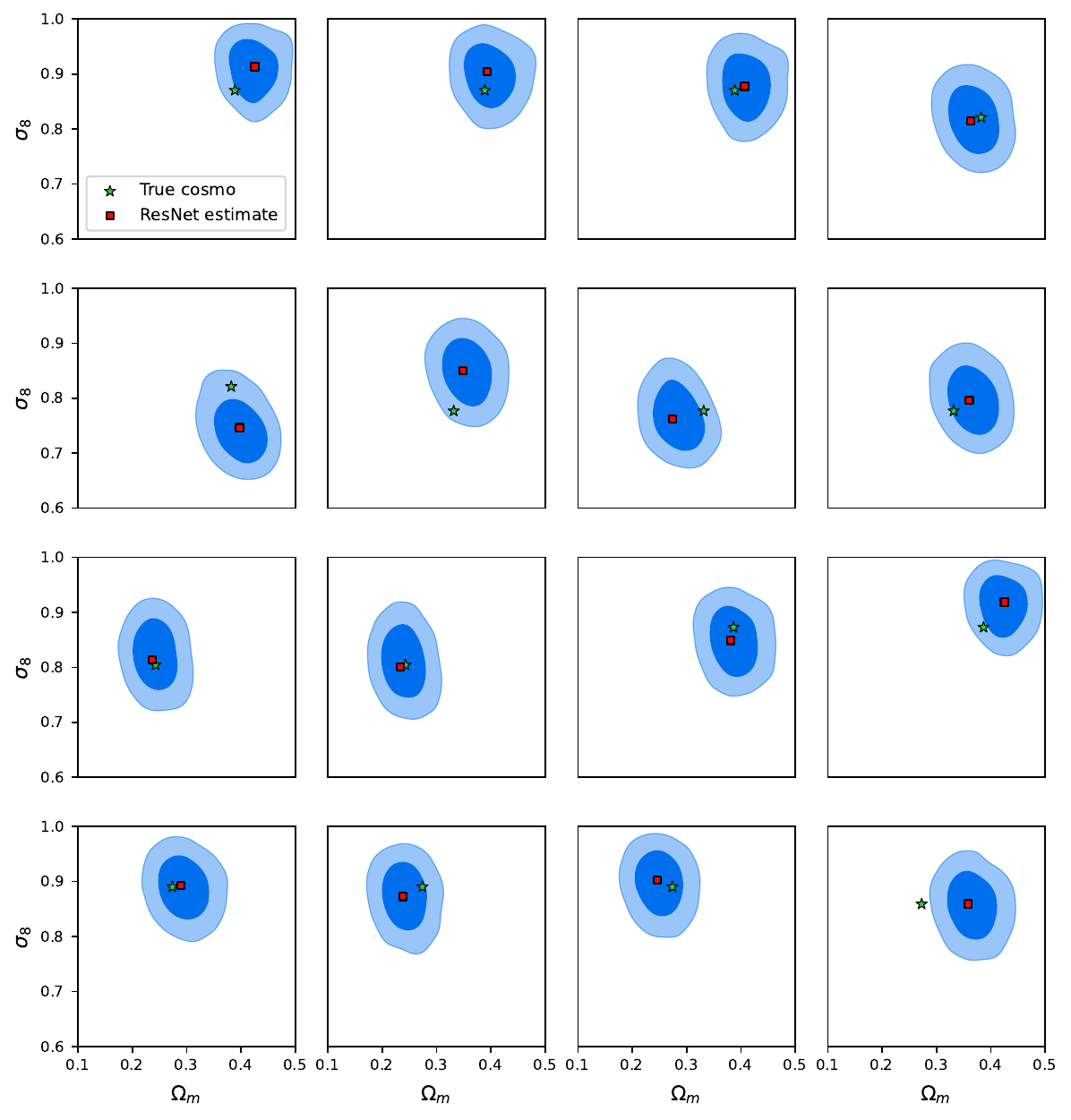}
    \caption{NPE performed at different points in the parameter space. XODs are produced from 200 deg$^2$ X-ray surveys, selecting clusters with the flux cut $CR_{lim}=0.02$c/s. The posteriors (blue contours) are obtained with the NPE for XODs unseen during the ResNet nor NPE trainings. A tested XOD $x_0$ is compressed into a $y_0$, which position in the parameter space is indicated by a red square. The green star shows the true parameters that were used to simulate $x_0$.}
    \label{fig:SICC_contours_manycosmo}
\end{figure*}

 Finally, we present the results of the NPE method. We use simulated XODs $x_0$ that were neither used for the ResNet training nor its testing, and use the ResNet to obtain their compressions $y_0$. The compressions are in turn passed to the trained NPE, providing us with a posterior estimations $q(\theta\mid y=y_0)$, marginalized over the feedback parameters. We show the 68\% and 95\% credibility regions on cosmology, marginalized over the feedback parameters, for 16 different $y_0$ in figure \ref{fig:SICC_contours_manycosmo}, where we have simply excluded XODs simulated from parameters close to the border of the parameter space. We show in addition the position of the compression $y_0$, and of the true underlying cosmological parameters $\theta_{true}$. We stress that all $\theta_{true}$ shown are different. We add in appendix \ref{app:SICC_rank} an analysis of the quality of our posteriors. This illustrates the viability of our model for a cosmological inference. The acceleration of the simulations with GPU resources and the LDL surrogate baryonification enables the creation of the datasets required by the SBI approach we use. 
\par
\begin{figure}
    \centering
    \includegraphics[width=0.9\linewidth]{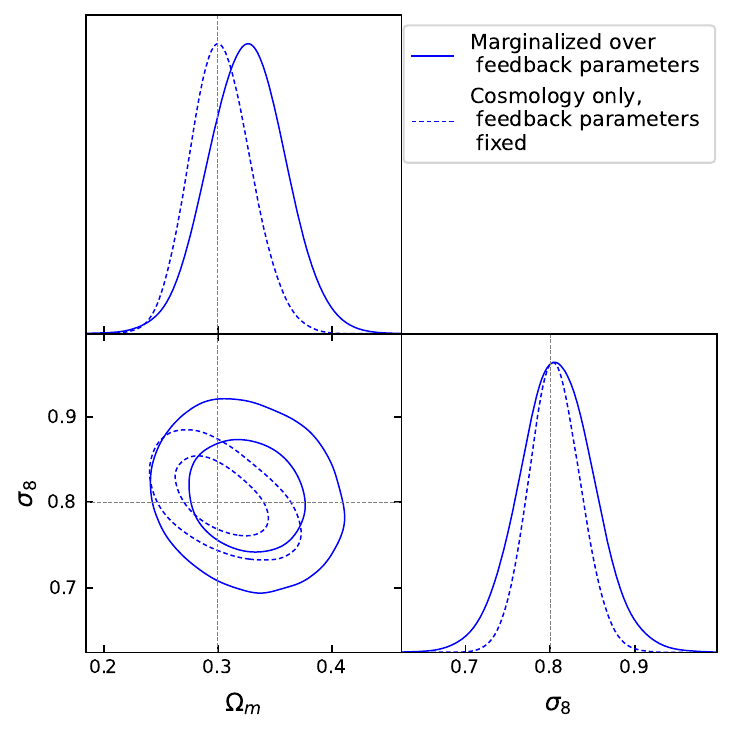}
    \caption{Posteriors estimated from a fiducial XOD for different conditioning of the NPE. The XOD is produced from a 200 deg$^2$ X-ray survey, selecting clusters with the flux cut $CR_{lim}=0.02$ c/s. In plain blue is shown a posterior marginalized on the feedback parameters, while in dashed blue is shown a posterior from a NPE only conditioned on the cosmology. The vertical and horizontal lines recall the fiducial values used for generating the tested XOD.}
    \label{fig:SICC_RDvsCO}
\end{figure}

 In figure \ref{fig:SICC_contours_manycosmo}, the commonly observed degeneracy between $\sigma_8$ and $\Omega_m$ in cluster counts experiments appears notably weak. While this degeneracy typically arises from the role that $\sigma_8$ and $\Omega_m$ play in shaping the HMF, the impact of nuisance parameters on the degeneracy may vary between different models of X-ray cluster properties. In our framework, we test this by comparing the inferred posteriors obtained with and without including feedback parameters in the analysis. For this purpose, we run a new set of simulations with fixed astrophysics, reuse our ResNet to convert the XODs into neural compressions, and finally retrain a posterior estimator conditioned solely on the cosmology. We take a fiducial diagram $x_0$, pass it to both posterior models (one conditioned on all parameters and one conditioned only on cosmology), and show the resulting contours in figure \ref{fig:SICC_RDvsCO}. Our model retrieves a more pronounced degeneracy between $\sigma_8$ and $\Omega_m$, with feedback parameters are fixed. Interestingly, including them in the inference broadens the posterior, in particular along the transverse direction to the degeneracy. We recall that feedback is also accounted for in a scaling relations-based model, but through an empirical parametrization that broadens the original $\Omega_m$-$\sigma_8$ contour in a different manner.

\section{Discussion}\label{sec:discussion}
\subsection{Intrinsic degeneracies of the simulation-based and analytical models}

To assess the relevance of our new modelling, we first study the intrinsic degeneracies of both models. For the analytical forward model, we can explicitly write the likelihood. Writing an XOD as a set of expected number counts $\lambda_i(\theta)$ in bins of the ($CR$, $HR$, $z$) space, and if we assume that observed counts in different bins are independent and follow a Poisson distribution, we can express the Fisher matrix in the following manner:
\begin{equation}
    F_{\alpha\beta} = \sum_i \frac{1}{\lambda_i}
                       \frac{\partial \lambda_i}{\partial \theta_\alpha} 
                       \frac{\partial \lambda_i}{\partial \theta_\beta}
.\end{equation}
Under the assumption that the likelihood is Gaussian, the Fisher matrix can hence be inverted to obtain the optimal constraints achievable by the model. More details on the computation of the Fisher matrix can be found in \cite{cerardi_cosmological_2024, kosiba_cosmological_2024}. 

\par
\begin{figure*}
  \centering
  \includegraphics[width=9.5cm]{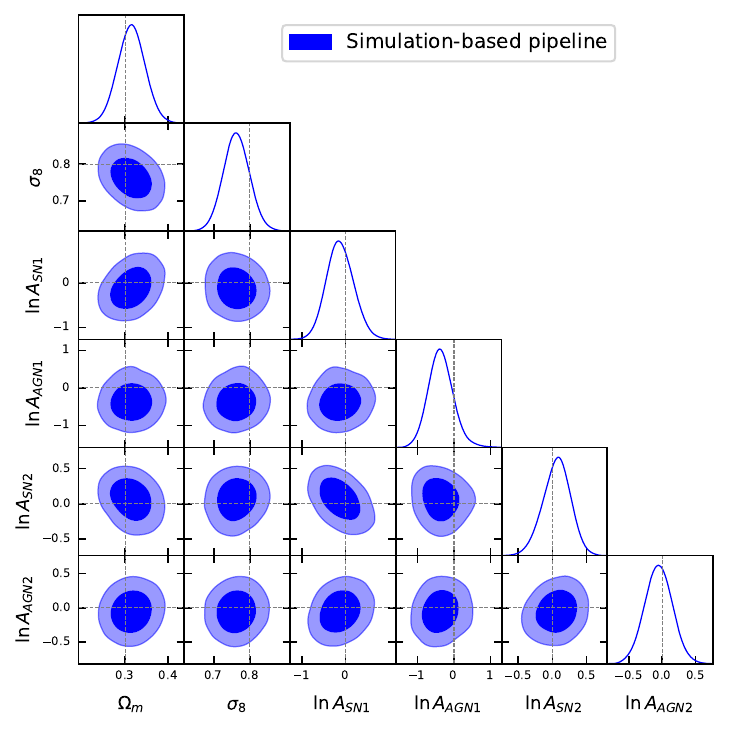}
    \caption{Posteriors on the cosmological and astrophysical parameters, using the simulation-based forward model and the NPE inference method. The tested diagram $x_0$ is drawn from the fiducial model. XODs are produced from 200 deg$^2$ X-ray surveys, selecting clusters with the flux cut $CR_{lim}=0.02$c/s. }
    \label{fig:SICC_posteriors_full}
\end{figure*}

\begin{figure*}
  \centering
  \includegraphics[width=12cm]{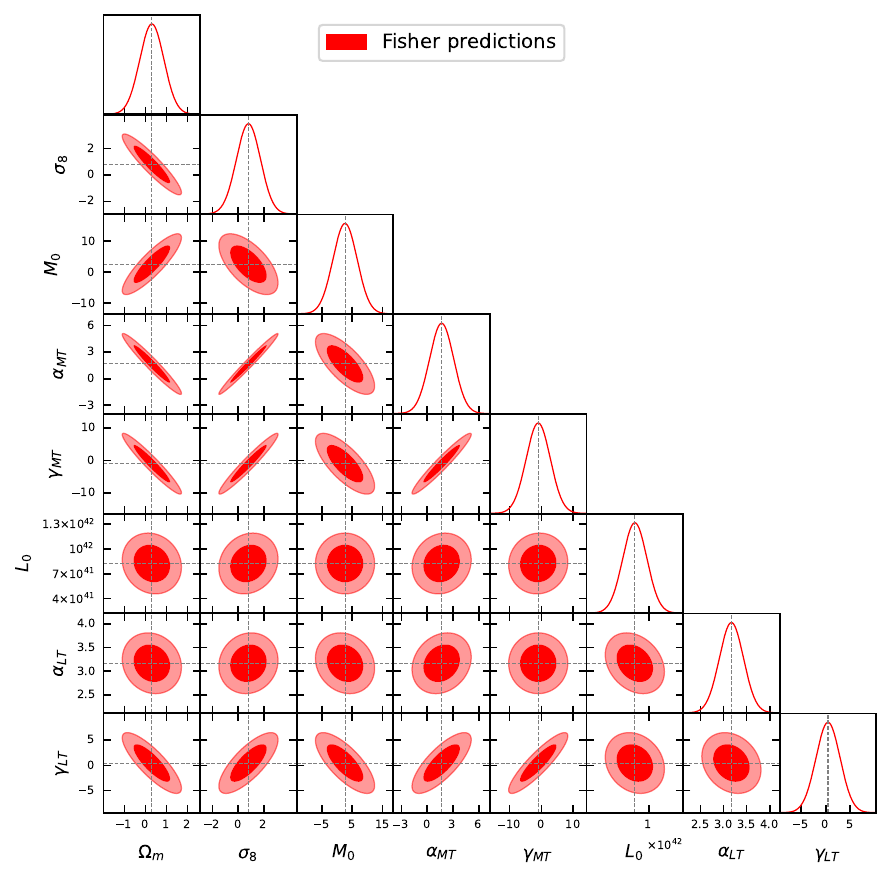}
    \caption{Posteriors on the cosmological and nuisance parameters, using the analytical forward model and the Fisher analysis, with explicit scaling relations. We observe that many  pairs of parameters are degenerate (elongated contours), denoting the ill-parametrization of this model. The simulation-based forward model does not show this behaviour, in figure \ref{fig:SICC_posteriors_full}. XODs are produced from 200 deg$^2$ X-ray surveys, selecting clusters with the flux cut $CR_{lim}=0.02$c/s.}
    \label{fig:fisher_posteriors}
\end{figure*}

 We show in figure \ref{fig:SICC_posteriors_full} and \ref{fig:fisher_posteriors} the full posteriors for the simulation-based and the analytical model, respectively obtained through the NPE and Fisher analysis methods. An elongated contour indicates a degeneracy between parameters. Many clear degeneracies appear in the model with scaling relations: not only the scaling relation coefficients are degenerate between each other (e.g. $\alpha_{MT}$ and $\gamma_{MT}$), but in addition, they are also degenerate with the cosmological parameters (e.g. $M_0$ and $\Omega_m$). This shows that even under our XOD statistic \citep[which is superior to a representation in the mass redshift plane, see][]{clerc_cosmological_2012}, these parameters are hard to separate.
We do not observe similar degeneracies for the simulation-based model. With the current set of parameters, only $A_{SN1}$ and $A_{SN2}$ seem slightly degenerate but, most importantly, none of the feedback parameter appears degenerate with cosmology. This is the consequence of our approach being based on feedback prescriptions describing the physical processes within the ICM. Our model carries inherent physical constraints on cluster observable properties. In this proof of concept, we assume that feedback models are correct (within the freedom given to the four feedback parameters), which introduces a lot of information in the inference. 

\subsection{Towards a universal parametrisation ?}
\begin{figure}
  \centering
  \includegraphics[width=8cm]{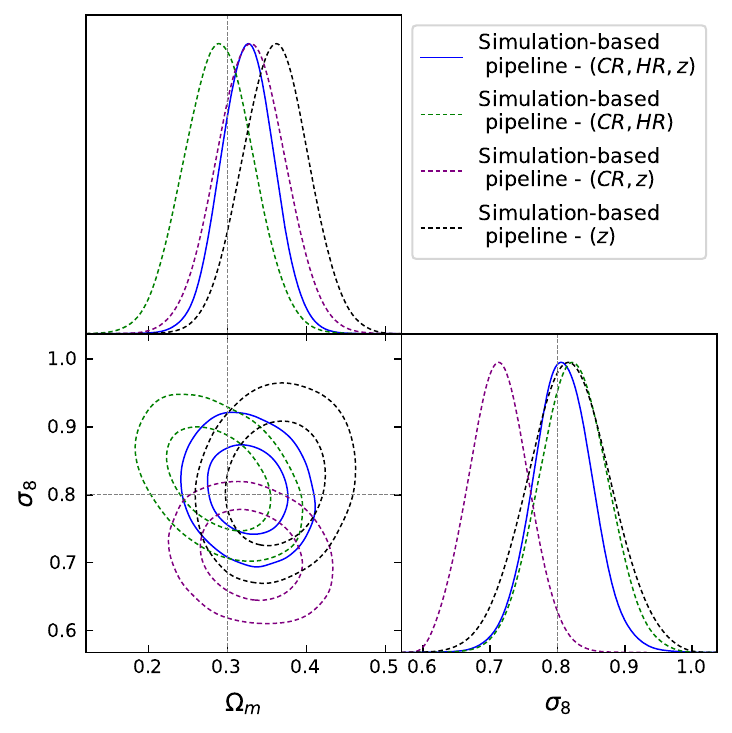}
    \caption{Cosmological posteriors obtained with the simulation-based forward model and the NPE. We trained the NPE for several statistics: the full ($CR$,$HR$,$z$) XOD (blue contours), the ($CR$,$HR$) number counts (dashed green), the ($CR$,$z$) number counts (dashed purple) and only the $z$ distribution (dashed black). The same simulations, forward model, and parametrisation are used, and we still recover informative constraints for all these cases.}
    \label{fig:SICC_posteriors_Oms8}
\end{figure}

The analytical modelling uses a set of empirical scaling relations that has to be well adapted to the chosen study: there is no general modelling available (see for instance the different references in table \ref{tab:compare_studies}). Depending on the observables and on the characteristics (more or less massive, more or less distant) of the detected sources, one would change the number of scaling relations used and the number of free parameters to best fit the data (see the compilation in table \ref{tab:compare_studies}). For instance, the eRASS1 analysis does not use the $HR$ (thus no $L-T$ or $M-T$ relation needed), but does include the optical richness $\lambda$ (so an extra relation $M-\lambda$ has to be modelled), and several model parameters have a redshift dependence at the cost of extra nuisance parameters. As the parameters of our simulation-based model are related to the physical processes in the ICM, we shall not need need to change our parametrisation for a different set of observables. We show in figure \ref{fig:SICC_posteriors_Oms8} the contours obtained if we run the analysis by simply removing one or two dimensions in the ($CR$, $HR$, $z$) space. The contours are slightly broadened because of the loss of information, but the constraints are still informative. This can also be seen in the table \ref{tab:sicc_sbi_diffstats} that compares the relative figure of merit of these posteriors. \cite{clerc_cosmological_2012} found that the redshift distribution struggles to constrain cosmology, because of degeneracies between parameters that the $CR$-$HR$ representation is able to break. With our simulation-based model trained on astrophysical processes, the dn/dz statistics provides constraints that are comparable to those from the $CR$-$HR$ space. This prediction is quite different from that by \cite{clerc_cosmological_2012} (fig. 12) and suggests, that because our set of (non-cosmological) priors is more physically motivated than theirs, our scaling-relation independent modelling of cluster properties is less degenerate. Our approach seems thus to ``naturally’' exclude entire unphysical regions of the coefficient-cosmology space, that were left open in the approach relying on scaling relations with priors. But, in this comparison, one has to consider that  \cite{clerc_cosmological_2012} implemented measurement errors in the CR-HR space and had one more free cosmological parameters ($w_{0}$), which opened additional degeneracies.

\begin{table}
\centering
\caption{Performances of different test statistics with the simulation-based model}\label{tab:sicc_sbi_diffstats}
\begin{tabular}{cccc}
\hline
\makecell{Observables used in \\the summary statistics} & FoM   &   $\Delta \Omega_m \ \left( \frac{\Delta \Omega_m}{\Omega_m}\right) $  & $\Delta \sigma_8  \ \left( \frac{\Delta \sigma_8}{\sigma_8}\right)  $    \\ \hline \hline
         $z$,$CR$,$HR$  & 17  & 0.068 (23\%)& 0.090 (11\%) \\
         $CR$,$HR$ & 14 & 0.076 (25\%) & 0.11 (14\%)\\
         $z$,$CR$ & 10 & 0.082 (27\%) & 0.12 (15\%) \\
         $z$ & 9 & 0.087 (29\%) & 0.13 (16\%) \\ \hline
\end{tabular}
\tablefoot{We obtain posteriors from different test statistics, and display the $68\%$ credibility intervals for $\Omega_m$ and $\sigma_8$, as well as the figure of merit ($\text{FoM}=\text{Cov}(\Omega_m, \sigma_8)^{-1}$). The same parametrisation is used for each  test statistics and informative constraints are retrieved. This highlights the universality of our physical modelling based on simulations.}
\end{table}

\subsection{Scaling relations conditioned on astrophysical parameters}

\begin{figure}
  \centering
  \includegraphics[width=8cm]{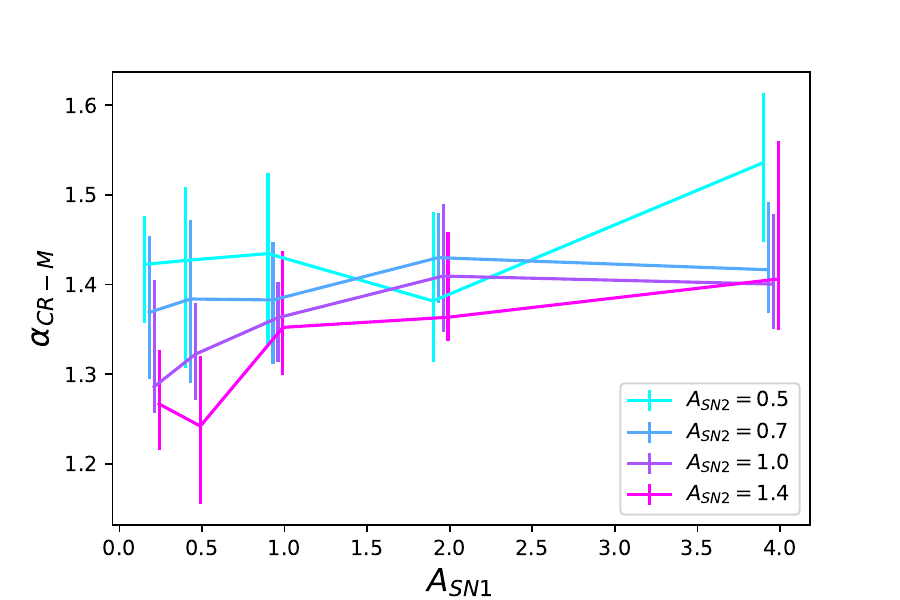}
  \includegraphics[width=8cm]{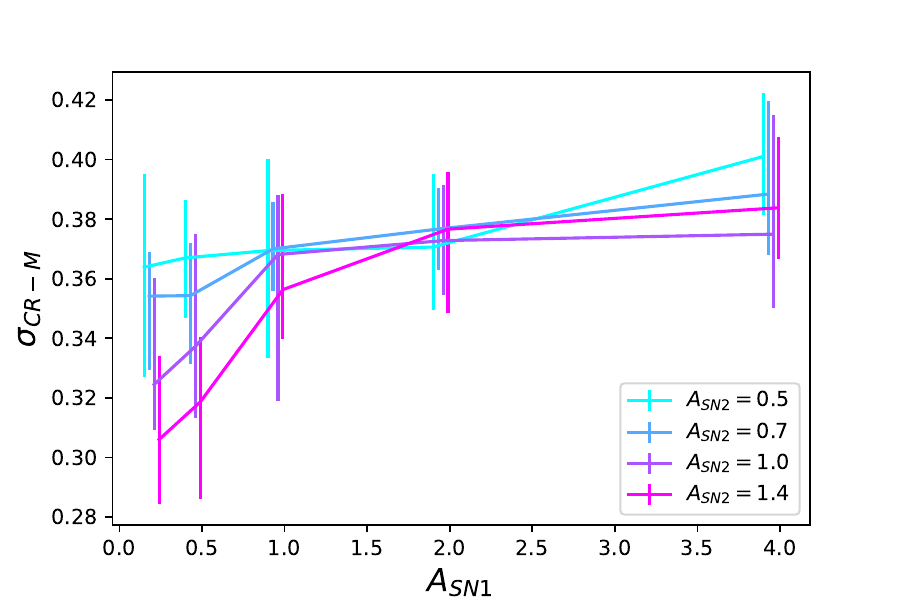}
   \caption{$CR$-$M$ relation coefficients as a function of the simulation parameters, for large volumes emulated at $z=0.21$. The $x$-axis varies the $A_{SN1}$ feedback parameter, while the line colour denotes a change in $A_{SN2}$. The top and bottom panel respectively show the variations of the slope $\alpha_{CR-M}$ and of the intrinsic scatter  $\sigma_{CR-M}$. The error bars show the interval covering 68\% of the measured point around the median. The lines have been slightly shifted horizontally to improve readability.}
    \label{fig:SL_abaque1D}
\end{figure}

In section \ref{sec:results_eLDL} we studied the $CR$-$M$ scaling relations emulated in our simulations under the fiducial set of parameters. This  opens a new way of modelling cluster number counts: we can use our simulation-based model to compute new scaling relation coefficients when changing the cosmological and the feedback parameters. In figure \ref{fig:SL_abaque1D}, we track the changes of the slope and scatter (two coefficients that have theoretical predictions, from which actual values may differ because of non gravitational processes) of the $CR$-$M$ law when changing $A_{SN1}$ and $A_{SN2}$. We simulated for each set of parameters 16 simulations of volume ($150\times150\times50$) $h^{-3}$Mpc$^3$, at $z=0.21$, projected on the last dimension. This represents 144 times the volume of one CAMELS simulation from the CV set, and hence provides us with many cluster detections to compute scaling relations. In the top panel of figure \ref{fig:SL_abaque1D}, showing the variations of $\alpha_{CR-M}$, the error bars are large hence only general trends can be retained. We see that the parameter slightly increases (resp. decreases) with increasing (decreasing) $A_{SN1}$ ($A_{SN2}$). This response of the slope parameter could explain the weak negative degeneracy observed in the full simulation-based posterior, figure \ref{fig:SICC_posteriors_full}, in the $A_{SN1}$-$A_{SN2}$ region. The bottom panel of figure \ref{fig:SL_abaque1D} shows a cross effect of the two SN feedback parameters on the  slope. At low values of $A_{SN1}$, $\sigma_{CR-M}$ is strongly affected when $A_{SN2}$ is varied. Interestingly, this behaviour does not appear for high $A_{SN1}$ values, at least for the current level of uncertainty on the data points. In appendix \ref{app:abaque_4D}, we provide a more complete view of the dependence of the scaling relations on the simulation parameters, including $\Omega_m$ and $\sigma_8$.
\par With a simulation-based model that considers all relevant astrophysical parameters from simulations, we could build an emulator of plausible scaling relations as a function of cosmological and astrophysical parameters and, subsequently, forward model cluster counts with these emulated scaling relations. In this way, we would sample the cosmological and astrophysical parameter space, avoiding forbidden regions and the  degenerate parametrisation of the scaling relations. This would also allow an important gain in speed, as XODs created with emulated scaling relations would require less computational resources than with the simulation-based model. Traditional inference techniques would be available for this type of forward model, as we would obtain theoretical number counts, instead of Poisson realisations of XODs. But we would loose the spatial information on the cluster properties within the cosmic network (not used in this paper).

\subsection{Towards an application on real data ?}\label{sec:discussion_realdata}
In order to apply our method for an inference with real data, several points should be carefully considered. The third question raised in the introduction, relative to the realism of hydrodynamical simulations, becomes here critical. We noticed in section \ref{sec:results_eLDL} that our simulation-based model appears to predict more clusters than the analytical method. We have ruled out a systematic error from the LDL prediction , as the luminosity mass relation for groups and clusters (figure \ref{fig:SICC_SL}) is well reproduced. Candidates for this discrepancy might be the X-ray properties emulated in the CAMELS/IllustrisTNG simulations. As a result of the CAMELS implementation, SN and AGN are tuned to provide realistic star formation histories, but not the overall properties of the X-ray cluster population. We hence highlight the importance of calibrating the sub-grid physical models on cluster-related observables. Beside the simulations used for calibration, another critical point is the detection method. It is important for a forward model to integrate the same biases than the ones inherent to the processing of real data. Given that our method produces mock X-ray observations, the straight forward solution is to apply the same detection pipeline on both the real and mock data. To do this, some simplifications taken here have to be dropped: it becomes necessary to include AGN contaminants, CCD defaults, realistic diffuse and proton backgrounds, and a Poisson noise on the pixels photon counts. Concerning the measurement, error models are needed for a scaling relation-based modelling \citep[e.g. in][]{garrel_xxl_2022}, but not in a simulation-based modelling. Indeed, if the mocks and the detection are fully coherent with the observed data, the measurement errors would be encoded in the simulation output. As a result, we would not need an external model for that, as we do for a classical analysis.
\par A second improvement is expected regarding the simulations that serve to train the extended LDL. \cite{villaescusa-navarro_camels_2021} has for instance found that the star formation rate density in the original CAMELS was not much sensitive to changes in the two selected AGN parameters varied (only for Illustris/TNG). In addition, we have also looked at the trend between $n_e$, $T$ and $\rho_{DM}/\rho_c$, at the voxel level in the CAMELS 1P set, which varies only one parameter at a time. We find that the gas properties are more affected by the SN parameters than by the AGN feedback. This could explain why our extended LDL is not very sensitive to these parameters. The surprising strong impact of SN feedback is in fact indirect: the coupling between SN and AGN feedback effectively tempers the latter, explaining the opposite trend from increased AGN activity \citep{tillman_exploration_2023, medlock_quantifying_2024}. New versions of CAMELS simulations are being run that include the variations of additional physical parameters. More AGN feedback parameters will be implemented, and specifically the threshold mass triggering the AGN kinetic feedback mode, a parameter that is known to significantly impact the X-ray luminosity of extragalactic gas \citep{truong_correlations_2021}. As a result, the extended LDL could be conditioned on parameters that effectively impact X-ray observables, and the learnt posterior could be marginalized on them. It will be necessary to test our approach under this new configuration to assess whether the additional parameters introduce degeneracies with the cosmology. 
\par Another important asset of this new CAMELS dataset is the increase of the simulated box size, firstly to (50$h^{-1}$Mpc)$^3$ and later to (100$h^{-1}$Mpc)$^3$. This firstly will form more clusters, and more massive ones, leading to more training samples for the LDL approach. Also, we have tested the fidelity of LDL emulated scaling relations only on the fiducial model, because the small boxes in the LH set only produce a handful of groups and lightweight clusters, and do not provide us with sufficient statistics to compute scaling relations. The larger boxes of the next CAMELS iteration will provide us with more statistics. It could represent for us an opportunity to analyse the LDL generated clusters for non fiducial sets of parameters.
In addition, training and testing the extended LDL on more massive clusters will improve the consistency of our method when applied to bright cluster samples.
\par Another needed improvement is relative to the computational resources. Our approach can only currently run on a single GPU, which limits drastically the number of particles in a simulation. Here, the limiting factor has been the simulated volumes for the lightcone creation. This has set the resolution of the LDL emulation, and hence we downgraded the CAMELS/IllustrisTNG simulations accordingly. An undergoing work (Kabalan et al., in prep) aims at parallelizing critical parts of our simulations, which will allow us in the end to improve the resolution of our X-ray mocks. This is of particular importance as cluster cores can exhibit very diverse emission characteristics, and because the matter distribution in clusters contains information on the accretion and merger history of a given cluster.

\section{Conclusions}

In this study, we have presented a novel approach to forward model cluster number counts in deep X-ray surveys. We used cosmological simulations, that integrate the dynamical and baryonic processes affecting the cluster luminosities. We have developed an accelerated baryonification technique by extending the LDL approach. The LDL uses particle displacements from corrections to the gravitational potential and non linear activation to transform a DMO field into a targeted baryonic property field. As such, it is motivated by physical arguments and simpler than deep generative models. We allow the baryonic properties to be conditioned on cosmology and on the feedback parameters. Moreover, our baryon pasting also depends on redshift. We adopted a novel, flexible Fourier filter to implement the LDL model and, subsequently, derived  the cluster properties down to X-ray observables, i.e. XMM countrates in several bands. We created a large number of  lightcones  ($0.1<z<0.5$, 200 deg$^2$) to perform  simulation-based cosmological inference for the  cluster population in the CR-HR-z parameter space. We compared our results  with the analytical approach. 
\par Our method not only avoids inference on cosmology-dependent measurements (such as mass and luminosity), but also bypasses the use of empirical scaling relations to model the X-ray cluster properties. We now examine its impact on the three key questions raised in the introduction:
\begin{itemize}
    \item \textbf{Accuracy of a fast ML emulator.} We have demonstrated that our extended LDL model was able to reproduce the cluster population from the original simulations, for the fiducial parameters. Moreover, our model does more than just recovering the statistical properties of the population, it also partially captures the specificity of each individual cluster. In that sense, our modelling is superior to an approach based on scaling relations. However, due to the lack of simulated volume for non-fiducial parameters, we cannot run the same test in the whole parameter space. Improvements in the emulator resolution could help fixing the remaining uncertainty on the X-ray fluxes at the individual cluster level. The availability of larger simulation boxes for some regions of the parameter space would help both the training and the testing of our model. The upcoming versions of CAMELS will satisfy this requirement. 
We recall that we chose a physics-based, lightweight ML approach. Conversely, one could try deep neural architecture to perform the baryonification, but at the cost of interpretability. It would be very informative to assess which approach performs better.
    \item \textbf{Viability for cosmological inference}. Our fast baryonification method reduces the computational cost of a simulation-based forward model and thus enables cosmological inference. Compared to the analytical modelling, our approach is less degenerate and allows us to infer nuisance parameters that have a more physical meaning than scaling relation coefficients. Our parametrisation, indeed,  is directly linked to well-identified astrophysical processes, and hence is in principle more universal: it is insensitive to changes in the survey design or in the chosen summary statistics. Nonetheless, we can only let free the parameters that are varied in the original CAMELS dataset, and it could be important to marginalize over other simulation parameters that are here fixed. Once again, the future developments of CAMELS will offer the opportunity to enlarge the conditioning of our extended LDL model.
    \item \textbf{Realism of the hydrodynamical simulations.} Although this is not the main issue targeted in this work, the 40\% difference between the simulation-based and analytical number counts may suggest a lack of realism. We could here question either the realism of the simulations, or the fiducial values chosen for the feedback parameters, or the implementation of the retroaction mechanisms themselves. However, the gap we observe could also arise from our simplified cluster detection and characterization process. Both, upgraded physics prescriptions in the hydrodynamical simulations and a more  realistic detection chain will be necessary to apply our method to observed cluster samples. Although the output resolution of the simulations is not a critical point for our goals, the resolution of the hydrodynamic solver may well play a critical role on the realisms of the simulations.
\end{itemize}
\par Further work will be dedicated to deeper analysis of the clusters emulated by the extended LDL, in order to charactereise their profiles. We also plan to investigate  de discrepancy  with the observational number counts by upgrading the production of the X-ray mocks. Regarding the fast baryon pasting approach, we will leverage the new large simulation sets in CAMELS to improve our analysis of the conditioning step. This will also allow us to more accurately map the scaling-relation coefficients as a function of the physical parameters (here SN and AGN feedback) as sketched in figure \ref{fig:SL_abaque1D}. Consequently, we shall  be able to identify regions of the cosmology-coefficient space that are a priori forbidden by cluster physics. This would provide a novel approach to setting priors on the coefficients. 
\par During the review process of this article, other works on SBI for cluster cosmology have been made available on arXiv \citep[][ respectively for SZ and X-ray cluster samples]{zubeldia_extracting_2025, regamey_galaxy_2025}. This demonstrates the growing interest in this new class of inference methods for cluster cosmology. However, these approaches differ from ours as they both employ scaling relations to forward model cluster properties.
\par Our simulation-based forward approach opens new avenues for implicit modelling and inference in cluster cosmology. In particular, when we are in a position to satisfactorily reproduce  the observables properties of each individual cluster, the X-ray mapping  will implicitly carry a lot of supplementary information relative to cluster environments and histories. 
This could in turn be implemented in the cosmological inference in terms of new observables. We, hence, can think of switching from  CR-HR-z  to  a full spatial mapping of the cluster population in the form of CR-HR-z-x-y (x = RA, y = Dec). This would allow us to combine the current XOD with an information comparable to the 2 pt-correlation function. But providing much more stringent constraints, since the X-ray cluster properties would be related to their location within the cosmic network.
In this way, we anticipate that the 5D new summary statistics would exhaust the cosmological potential of an X-ray cluster survey, and remove most of the degeneracy in cluster cosmology. Joint modelling of other cosmological probes along with clusters is also a possibility within this framework, and promises to break degeneracies between nuisance and cosmological parameters \citep{omori_agora_2022}.

\begin{acknowledgements}
      The authors thank Camila Correa, Jean Ballet, Nicolas Clerc and Daisuke Nagai for relevant discussions around this work. The authors also thank the referee for insightful comments on this work. This work was granted access to the HPC resources of IDRIS under the allocations 2023-AD011013487 and 2024-AD011013487R2 made by GENCI. NC is grateful to the IDRIS technical team for their support on the Jean-Zay computational resources. NC thanks Shy Genel and Francisco Villaescusa-Navarro for sharing the CAMELS CV50 simulations. NC thanks Marine Lafitte for graphical help on the figures.
\end{acknowledgements}

%
%
\bibliographystyle{aa}
\bibliography{LDLCosmo}

\begin{appendix} 
\section{Distribution}\label{app:LH_sampling}
\begin{figure*}
    \includegraphics[width=\textwidth]{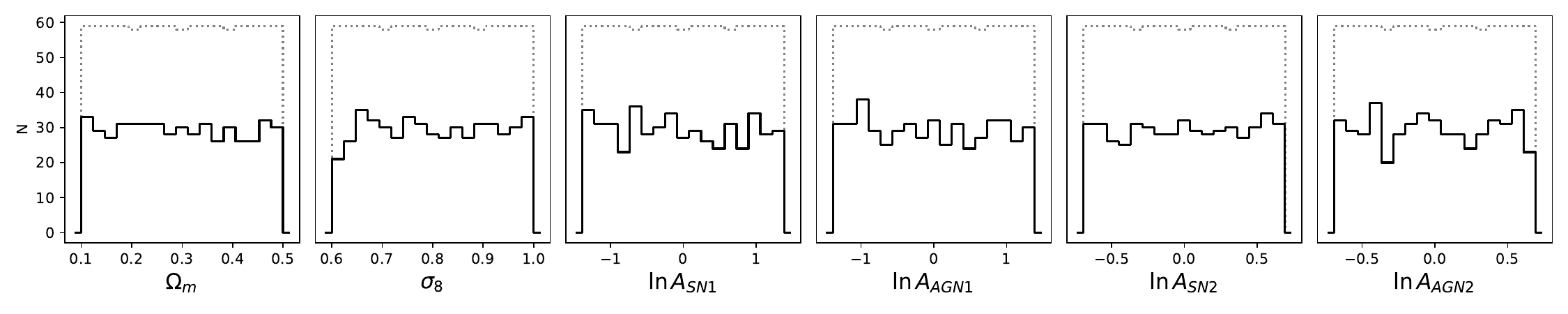}
    \caption{1D distributions of the simulations parameters in the CAMELS/LH suite. Dotted lines indicate the full suite (1000 boxes), while the plain lines show the actual distribution of the subset we used (499 boxes).}
    \label{fig:LH500_sampling}
\end{figure*}

We use the CAMELS simulation suite to train and test our extended LDL method. In figure \ref{fig:LH500_sampling}, we show the parameter distribution for each of six varied parameters, in the CAMELS/LH set. For completeness, we show both the distribution for the full suite (dotted lines, 1000 samples), and for the subset we used (plain lines, 499 samples). We do not observe a specific concentration of the simulated boxes parameters in the 1D distributions. By construction, the LH set samples parameters mimicking a uniform (or log-uniform for the feedback parameters) distribution. Our subset, even if it divides the density of points by a factor of 2, still conserves a rather regular sampling.

\section{Quality of the posteriors}\label{app:SICC_rank}
\begin{figure}
\includegraphics[width=8cm]{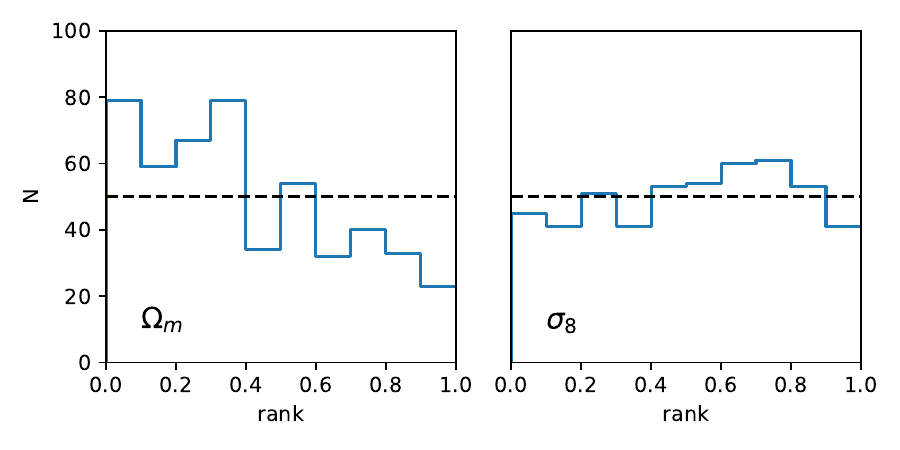}
\caption{Ranks of the true $\Omega_m$ (left) and $\sigma_8$ (right) in the 1D marginalized posteriors. We have tested 500 different compressions $y_0$, leading to the distributions represented in the blue histograms. If the posteriors are unbiased and with the proper width, the observed distribution should be pproximately flat (dashed black line). This is the case for $\sigma_8$. However, $\Omega_m$ exhibits a negative slope that indicate a bias, but still an appropriate posterior size (no apparent curvature).}
\label{fig:SICC_ranks}
\end{figure}

We here assess the quality of the 1D posteriors on the cosmology. We begin with 500 XODs $x_0$ unseen during the ResNet and NPE trainings, which we compress into $y_0$. We followingly obtain 500 6-dimensional posteriors $q_{\varphi}(\theta\mid y=y_0)$. We then obtain the 1D posteriors over $\Omega_m$ and $\sigma_8$ by marginalising over the other parameters. We compute the 1D rank for both $\Omega_m$ and $\sigma_8$:
\begin{equation}
    r(\theta_{true},y_0)=\int_{\theta_{min}}^{\theta_{true}}p(\theta\mid y=y_0)\diff \theta,
\end{equation}
with $p$ the 1D marginalised posterior. We then represent the distribution of $r(\theta_{true},y_0)$, for both parameters, in figure \ref{fig:SICC_ranks}. A flat trend indicates no bias, while a slope reveals a bias. A concavity (resp. convexity) denotes an underconfidence (overconfidence) of the posterior. The linear trend for each plot indicates that our posteriors have the right size. However, the negative slope for $\Omega_m$ shows that we are biased on this parameters, which could be explained by the MSE loss in the regression. \cite{lanzieri_optimal_2024} have for instance found that this loss function can induce a biased posterior. We devote to a further study the exploration of other SBI methods within our framework.

\section{Deeper insights into the $CR-M$ scaling relation coefficients as a function of simulation parameters}\label{app:abaque_4D}

\begin{figure}
    \centering
    \includegraphics[width=8cm]{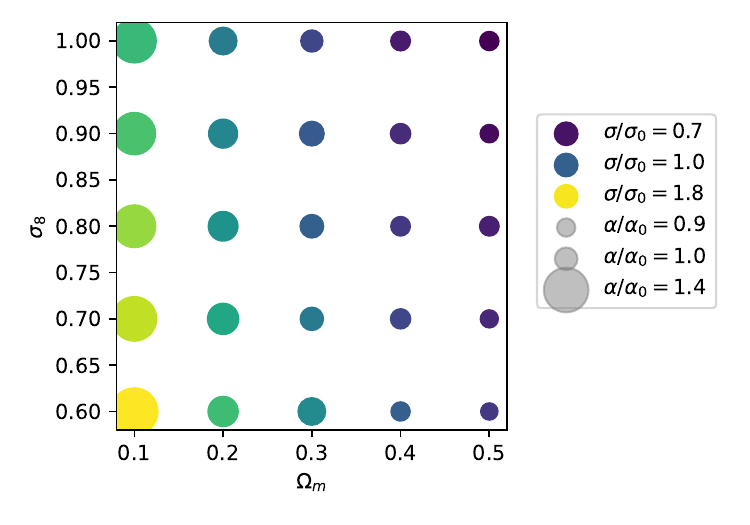}
    \includegraphics[width=8cm]{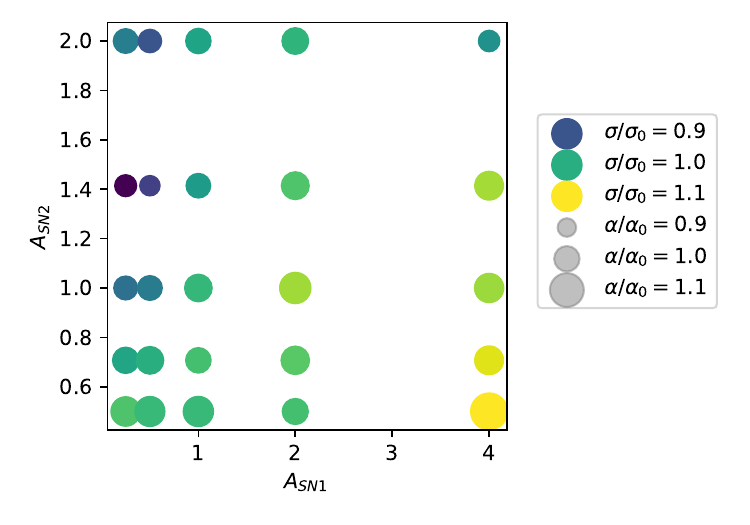}
    \caption{4D view of the dependance of the scaling relation coefficients on the simulation parameters. The marker size denotes the value of the scaling relation slope, and the marker color represents its dispersion. Both quantities are normalized with respect to the scaling relation of the fiducial model (central point).  
    \newline Top: Variation of the $CR$-$M$ slope and scatter when varying $\Omega_m$, $\sigma_8$, sampled on a linear grid.
    \newline Bottom: Variation of the $CR$-$M$ slope and scatter when varying $A_{SN1}$ and $A_{SN2}$, sampled on a logarithmic grid.}
    \label{fig:abaque4D}
\end{figure}
In figure \ref{fig:abaque4D}, we provide a 4D view of the scaling relation coefficients dependance on simulation parameters. In the $\Omega_m$-$\sigma_8$ plane (top panel), both the scatter and the slope of the $CR$-$M$ relation vary smoothly, with variations from -30\% to +80\% (resp. -10\% to +40\%) for the scatter (the slope). In the $A_{SN1}$-$A_{SN2}$ plane (bottom panel), we also observe trends, although the variations appear to be weaker ($\pm$10\% in all cases) and a bit noisier.
\end{appendix}

\end{document}